\documentclass[useAMS,usenatbib]{mn2e}
\usepackage{epsfig}
\usepackage{color} 
\usepackage{natbib}
\usepackage{threeparttable}
\usepackage{deluxetable}
\usepackage{amssymb}
\usepackage{rotating}
\usepackage{graphicx}
\usepackage[colorlinks=true,citecolor=blue]{hyperref}
\usepackage{mathpazo}
\usepackage{graphicx}
\usepackage{footnote}
\usepackage{longtable}

\title[INOV of X-ray versus $\gamma$-ray detected NLSy1s]
{Comparative intra-night optical variability of X-ray and $\gamma$-ray detected 
narrow-line Seyfert 1 galaxies}
\author[Ojha et al. ]{Vineet Ojha$^{1}$\thanks{E-mail: vineet@aries.res.in
  }, Hum Chand$^{1,~2}$, Gopal-Krishna$^{3,~1}$, Sapna Mishra$^{1}$, Krishan Chand$^{1}$   \\
$^{1}$Aryabhatta Research Institute of Observational Sciences (ARIES), Manora Peak, Nainital 
  $-$ 263002, India\\
$^{2}$Department of Physics and Astronomical Sciences, Central University of Himachal Pradesh (CUHP), Dharamshala$-$176215, India\\
$^{3}$UM-DAE Centre for Excellence in Basic Sciences, Univ. of Mumbai Campus, Mumbai$-$400098, India\\}

\begin{document}
\date{Accepted ---. Received ---; in original form ---}

\pagerange{\pageref{firstpage}--\pageref{lastpage}} \pubyear{2017}

\maketitle

\label{firstpage}
\begin{abstract}
In a systematic program to characterise the intra-night optical
variability (INOV) of different classes of narrow-line Seyfert 1
galaxies (NLSy1s), we report here the first comparative INOV study of 
NLSy1 sets detected in the X-ray and $\gamma$-ray bands. Our
sample consists of 18 sources detected in X-ray but not in
$\gamma$-rays (hereafter x$\textunderscore$NLSy1s) and 7 sources detected in $\gamma$-rays
(hereafter g$\textunderscore$NLSy1s), out of which 5
are detected also in X-rays. We have monitored these two sets of
NLSy1s, respectively, in 24 and 21 sessions of a minimum of 3 hours
duration each.  The INOV duty cycles for these two  sets are
found to be 12\% and 53\%, respectively (at a 99\% confidence
level). In the set of 18 x$\textunderscore$NLSy1s, INOV duty
cycle is found to be zero for the 13 radio-quiet members
(monitored in 14 sessions) and  43\% for the 5 radio-loud members
(10 sessions). The latter is very similar to the aforementioned duty
cycle  of 53\% found here for  the set of
g$\textunderscore$NLSy1s (all of which are radio-loud).  Thus it
  appears that the radio loudness level is the prime factor behind the
  INOV detection and the pattern of the high-energy radiation plays
  only a minor role.

\end{abstract}



\section{Introduction}
\label{sec1.0}

The unexpected discovery of $\gamma$-ray emissions from Narrow-line
Seyfert 1 (NLSy1) galaxies~\citep{Abdo2009ApJ...699..976A, Abdo2009ApJ...707..727A, 
Abdo2009ApJ...707L.142A, Foschini2010ASPC..427..243F, Foschini2011nlsg.confE..24F, 
D'Ammando2012MNRAS.426..317D, D'Ammando2015MNRAS.452..520D, Yao2015MNRAS.454L..16Y, Yang2018MNRAS.477.5127Y, Paliya2018ApJ...853L...2P,Yao2019MNRAS.487L..40Y}
and the realisation that essentially all of them have detection ($>3\sigma$) in the radio band too~\citep{Komossa2006AJ....132..531K, Yuan2008ApJ...685..801Y, Yao2015MNRAS.454L..16Y, Kynoch2018MNRAS.475..404K, 2018A&A...614L...1L}, suggesting a possible relationship between some NLSy1s and blazars.
 This hypothesis is further corroborated by the finding that the Spectral Energy Distributions (SEDs) of some NLSy1 galaxies show the canonical double-humped profile such that the lower energy hump representing the relativistic jet's
synchrotron emissions peaks somewhere between the near-infrared 
and X-ray frequencies.
Among blazars, such a wide range in the
synchrotron hump's peak frequency ($\nu^{p}_{syn}$) is observed only for their
subset called BL Lac objects. BL Lacs having $\nu^{p}_{syn}$
in the X-ray band are called HBLs and they are preferentially
picked up in X-ray surveys.
In contrast, BL Lacs with $\nu^{p}_{syn}$
falling in the near-infrared/optical part of SED are called LBLs
and they form a vast majority of the BL Lacs found in radio
surveys~\citep{Urry1995PASP..107..803U}.
For almost two decades it is known 
that, in comparison 
to LBLs, HBLs exhibit distinctly milder optical flux variability, both
on hour-like~\citep{Heidt1998A&A...329..853H, Romero1999A&AS..135..477R} and longer
time scales~\citep{Jannuzi1993ApJS...85..265J, Jannuzi1994ApJ...428..130J, Xie1996AJ....111.1065X}; 
the former being termed as Intra-Night Optical
Variability (INOV)~\citep{1995MNRAS.274..701G}. Further, it is found that the difference in INOV
persists even when only the TeV detected subsets of HBLs
are considered~\citep{Gopal-Krishna2011MNRAS.416..101G}. Since the entire synchrotron
hump of the AGN SED has normally been attributed to the same population 
of radiating relativistic electrons, the difference between the
optical variability of LBLs and HBLs has sometimes been attributed to the magnetic
field, by postulating a stronger field in the jets of HBLs, which would
tend to suppress the plasma turbulence in them, leading to 
their milder INOV~\citep{Sambruna1995ApJ...449..567S, Romero1999A&AS..135..477R}. 
However, the real situation could be more complex.\par

From the SWIFT near-infrared imaging of jets of the blazars 3C 273 and PKS
1136$-$135, it has been inferred that the optical emission component of the
SED is synchrotron emission arising from two distinct populations
of relativistic particles, one linked to the radio emission and the
other to the X-ray emission, the latter component could even be dominated by contribution from
relativistic protons~\citep{Jester2006ApJ...648..900J, Uchiyama2006ApJ...648..910U, 
Cara2013ApJ...773..186C}. This raises the
possibility that HBLs could differ from LBLs in a more fundamental
way and, if so, then the physical difference may even be mirrored in NLSy1 galaxies. 
Assuming that a possible X-ray linked optical synchrotron component is
more likely to provide a substantial, if not a dominant, contribution
in an X-ray detected sample of NLSy1s, a possible way to assess the
competing physical scenarios reported above is by comparing the INOV
characteristics of different NLSy1 samples exhibiting different high
energy properties; this is the approach we propose in the present
work.\par

NLSy1 galaxies are characterised by the narrow width of optical Balmer
emission line, with FWHM(H${\beta}$)$<$ 2000 km
s$^{-1}$~\citep{Osterbrock1985ApJ...297..166O, Goodrich1989ApJ...342..908G} 
and a flux ratio [O$_{III}]_{\lambda5007}/H\beta<$
3~\citep{Shuder-Osterbrock1981ApJ...250...55S}. Moreover, with some
possible exceptions, they exhibit strong [Fe VII] and [Fe X] line 
emissions~\citep{Pogge2011nlsg.confE...2P} and strong permitted lines 
of the optical/UV Fe~{\sc ii} emission~\citep{Boroson1992ApJS...80..109B,
Grupe1999A&A...350..805G}. They also show a steep soft X-ray spectrum
\citep{Boller1996A&A...305...53B, Wang1996A&A...309...81W, Grupe1998A&A...330...25G}, 
rapid X-ray and optical flux variability~\citep{Leighly1999ApJS..125..297L,
Komossa-Meerschweinchen2000A&A...354..411K, Miller2000NewAR..44..539M, 
Gaskell-Hedrick2004ApJ...609...69K}, and a frequently observed blue-shifted 
line profile~\citep{Zamanov2002ApJ...576L...9Z, Leighly2004ApJ...611..107L, Boroson2005AJ....130..381B}. Observations 
suggest that, compared to radio galaxies, NLSy1 galaxies tend to possess less 
massive central black holes and higher Eddington ratios (defined as the ratio 
of bolometric-to-Eddington luminosities, R$_{Edd}\equiv
L_{bol}/L_{Edd}\simeq1$) as compared to broad-line Seyfert galaxies (BLSy1s) 
and radio-quiet QSOs~\citep[e.g. see,][]{Boroson1992ApJS...80..109B,
Pounds1995MNRAS.277L...5P, Sulentic2000ApJ...536L...5S, 
Boroson2002ApJ...565...78B, Collin2004A&A...426..797C}. In addition, NLSy1 galaxies 
exhibit the radio-loud/radio-quiet bimodality displayed by QSOs and are also mostly radio-quiet~\citep[][and 
references therein]{Kellermann2016ApJ...831..168K}.  
The fraction of radio-loud NLSy1s (RL-NLSy1s), with the radio-loudness 
parameter\footnote{Radio-loudness is usually parametrised by the ratio
(R) of the rest frame flux densities at 5 GHz and at 4400\AA, being
R$\leq$ 10 and $>$ 10 for radio-quiet and radio-loud quasars,
respectively~\citep[e.g. see,][]{Visnovsky1992ApJ...391..560V, Stocke1992ApJ...396..487S, 
Kellermann1994AJ....108.1163K, Kellermann1989AJ.....98.1195K}.}
R$_{5GHz}  >  $ 10 is only $\sim$ 7\% ~\citep{Komossa2006AJ....132..531K, 
  Zhou2006ApJS..166..128Z, Rakshit2017ApJS..229...39R, Singh2018MNRAS.480.1796S},
 dropping further to just 2-3 percent for R$_{5GHz} >$ 100  
~(\citealp[see also] {Komossa2006AJ....132..531K}, \citealp[]{Zhou2002ChJAA...2..501Z}; 
\citealp[]{Yuan2008ApJ...685..801Y}). Comprehensive information on radio 
properties of NLSy1 galaxies can be found in~\citet{Lister2018rnls.confE..22L}.\par
\citet{Liu2010ApJ...715L.113L} reported the first detection of INOV in a $\gamma$-ray 
detected NLSy1, which showed that the NLSy1 galaxy J094857.30$+$002224.0 
exhibits INOV amplitude as large as 0.5 mag on a timescale of several hours. 
Similarly, for 3 {\it Fermi}-Large Area Telescope 
({\it Fermi}-LAT)\footnote{https://heasarc.gsfc.nasa.gov/docs/heasarc/missions/fermi.html} 
$\gamma$-ray detected NLSy1s, viz., J032441.20$+$341045.0,
J094857.30$+$002224.0 and J150506.48$+$032630.8,~\citet{Paliya2013MNRAS.428.2450P} have 
reported INOV with amplitude $>$ 3\% within a few hours, occurring with a very high duty cycle 
of $\sim$85\%. Recently, ~\citet{Kshama2017MNRAS.466.2679K} have compared the INOV 
properties of different classes of NLSy1s, consisting of $\gamma$-ray-loud NLSy1s 
(GL-NLSy1s, 4 sources), $\gamma$-ray-quiet NLSy1s (GQ-NLSy1s, 4 sources) and 
radio-quiet NLSy1s (RQ-NLSy1s, 4 sources). They found the INOV duty cycles (DCs) 
to be, respectively, $\sim$ 52\% and  $\sim$ 39\% for GL-NLSy1s and GQ-NLSy1s, while none 
of the RQ-NLSy1s showed INOV even at a modest (95\%) confidence level. We note that among 
their GL-NLSy1s and GQ-NLSy1s, a large fraction (viz., 100\% and 50\%, respectively) 
is radio-loud, supporting the inference that radio loudness is the hallmark of 
strong INOV in these AGNs.\par

On the other hand, a recent study of three $\gamma$-ray detected NLSy1s 
by~\citet{Ojha2019MNRAS.483.3036O} has indicated that superluminal motion in the 
radio-jet might be a more robust diagnostic of INOV. Therefore, it would be interesting to examine among
NLSy1s, whether the INOV has any link to other observed properties besides the radio-loudness and 
superluminal motion, such as high-energy radiation in X-ray and/or $\gamma$-ray bands.
 This is made possible by the fact that many NLSy1s have now been detected in the X-ray and/or $\gamma$-ray~bands, thanks to the various X-ray missions and the {\it Fermi}-LAT. Starting from the compilation of 76 NLSy1s by~\citet{Foschini2011nlsg.confE..24F}, we have carried out a study of the INOV characteristics of 18 NLSy1s detected in X-rays but not in  $\gamma$-rays (five of which are radio-loud), and 7 NLSy1 galaxies currently known to have a  $\gamma$-rays detection (all of which are found to be radio-loud).\par

The paper is organised as follows. In Sect.~\ref{section 2.0} we present an outline of our sample
of NLSy1 galaxies. Sect.~\ref{section3.0} describes their intranight optical monitoring and the 
data reduction procedure. Details of the statistical analysis are provided in Sect.~\ref{sec4.0}, 
while our main results followed by a brief discussion are presented in Sect.~\ref{section5.0}.

\begin{table}
  \begin{minipage}{80mm}
 \caption{The present sample consisting of 18 x$\textunderscore$NLSy1s (i.e., detected in X-rays but 
  not in ${\gamma}$-rays) and 7 g$\textunderscore$NLSy1s (${\gamma}$-ray detected).
  \label{tab:source_info}}
\begin{tabular}{rccr}
 \hline \multicolumn{1}{c} {SDSS Name{\footnote{Monitored with the 1.04m  Sampurnanand 
Telescope (ST) and also with the 1.3m DFOT ($^{\star}$). 
Monitored with the 1.04m ST only ($^{\dag}$).
The remaining sources were monitored using 
the 1.3m DFOT alone.}}}  & B-mag{\footnote{ Optical B-band magnitudes are taken from~\citet{Foschini2011nlsg.confE..24F}, except for two g$\textunderscore$NLSy1s, viz.,
J122222.99$+$041315.9 and J164442.53$+$261913.3, for which the USNO-A2.0 catalog was used~\citep{Monet1998AAS...19312003M}.}} & z{\footnote{ Emission redshifts values are taken from~\citet{Foschini2011nlsg.confE..24F}, except for the g$\textunderscore$NLSy1s, viz., 
 J122222.99$+$041315.9 and J164442.53$+$261913.3, for which the Sloan Digital Sky Survey release 10~\citep[SDSS DR-10,][]{Ahn2014ApJS..211...17A} catalogue was used.}} &
 $R_{1.4 GHz}${\footnote{$R_{1.4 GHz}\equiv f_{1.4 GHz}/f_{4400\AA}$ values are taken from~\citet{Foschini2011nlsg.confE..24F}, except
for J164442.53$+$261913.3 for which the value is taken from~\citet{Yuan2008ApJ...685..801Y} and 
J122222.99$+$041315.9 for which $R_{1.4 GHz}$ is estimated using a core flux density of 0.6 Jy 
at 1.4 GHz~\citep{Yuan2008ApJ...685..801Y}. The sources with $R_{1.4 GHz} > 19$ are termed here as radio-loud (e.g., see Sect.~\ref{section 2.0}). } } \\
\multicolumn{1}{c}{(1)}    &   \multicolumn{1}{c}{(2)}   &    \multicolumn{1}{c}{(3)}      &   \multicolumn{1}{c}{(4)}            \\
\hline
\multicolumn{4}{  c  }{\it  x$\textunderscore$NLSy1s}\\ \hline
J071340.30$+$382039.8    & 15.10 & 0.123  &  20  \\               
J073623.14$+$392617.9    & 16.88 & 0.118  &   3   \\        
J075245.60$+$261735.9    & 17.06 & 0.082  &      2   \\        
$^{\star}$J080638.98$+$724820.5    & 16.50 & 0.098  &    41    \\      
J093703.02$+$361537.1    & 17.99 & 0.180         &    12  \\
J100541.85$+$433240.2    & 16.87 & 0.179         &      4   \\
$^{\star}$J101000.70$+$300321.6    & 17.23 & 0.256  &      2   \\   
J140516.22$+$255533.9    & 15.46 & 0.165  &   1   \\       
$^{\dag}$J140827.81$+$240924.8    & 16.96 & 0.130  &   4    \\       
J144240.80$+$262332.6    & 17.06 & 0.108  &   5     \\
J144825.10$+$355946.7    & 16.87 & 0.114  &   2     \\
J151936.15$+$283827.6    & 17.34 & 0.270  &   4     \\ 
J162901.32$+$400759.5    & 18.04 & 0.272  &  50    \\
$^{\star}$J163323.59$+$471859.0    & 17.55 & 0.116  & 154   \\
$^{\dag}$J170231.05$+$324719.7  & 16.15 & 0.164  &   1   \\
$^{\star}$J170330.38$+$454047.3    & 16.52 & 0.060  & 102    \\
J171304.46$+$352333.4    & 16.56 & 0.085  &  10    \\
J171601.94$+$311213.7    & 16.18 & 0.110  &   1     \\
\hline
\multicolumn{4}{  c  }{\it  g$\textunderscore$NLSy1s} \\\hline
J032441.20$+$341045.0    & 16.38 & 0.063   &  318  \\
J084957.98$+$510829.0    & 19.27 & 0.584   & 4496 \\
J094857.32$+$002225.6    & 18.86 & 0.585   &  846  \\
J110223.37$+$223920.5    & 19.55 & 0.455   &   32   \\
J122222.99$+$041315.9    & 17.88 & 0.966   & 1534 \\
J150506.48$+$032630.8    & 18.99 & 0.408   & 3364 \\
J164442.53$+$261913.3    & 18.80 & 0.144   &  447  \\
\hline
\end{tabular}
\end{minipage}
\end{table}

\section{The sample}
\label{section 2.0}
Our sample for intranight monitoring has been drawn from the multi-wavelength compilation
by~\citet{Foschini2011nlsg.confE..24F} of 76 NLSy1s with
observations at high energies, namely X-ray (using
ROSAT\footnote{https://heasarc.gsfc.nasa.gov/docs/rosat/rosat.html})
and/or $\gamma$-rays (using {\it Fermi}/LAT). For each member of their
sample, these authors have estimated the radio loudness
parameter R$_{1.4 GHz}$, where R$_{1.4 GHz}$ is the ratio of the
monochromatic rest-frame flux densities at 1.4 GHz and
4400\AA~\citep[see][]{Yuan2008ApJ...685..801Y}. Note that the usual
dividing line of R = 10 between RQ-QSOs and RL- QSOs
~\citep[e.g.,][]{Kellermann1989AJ.....98.1195K} is defined for 5 GHz
and it corresponds to R$_{1.4 GHz}$ = 19 at 1.4 GHz, taking both radio
and optical spectral indices to be $-$0.5.\par
From observational considerations, we discarded 3 of the 76 NLSy1s which lie at
declinations south of $-30^{\circ}$.  Out of the remaining 73 NLSy1s,
18 have more than $3\sigma$ detection in X-rays (albeit undetected in
$\gamma$-rays), as well as B-band magnitude m$_{B}$ $\leq$ 18, hence
bright enough for intranight monitoring with the available 1-meter
class optical telescopes.
Five of the 73 NLSy1s have $> 3\sigma$ detection in $\gamma$-rays, among those, three have been detected in the X-rays too.
We have not imposed any optical brightness
threshold on these 5 NLSy1s, in view of the paucity of $\gamma$-ray
detected NLSy1 galaxies at present.  For the same reason, we have
included another two known $\gamma$-ray detected NLSy1s taken from the
literature, namely J164442.53+261913.3~\citep{D'Ammando2015MNRAS.452..520D} and
J122222.99+041315.9~\citep{Yao2015MNRAS.454L..16Y}. Thus, our sample
for INOV monitoring consists of 18 NLSy1 galaxies detected in X-rays
(but not in $\gamma$-rays) and 7 NLSy1 galaxies detected in
$\gamma$-rays (including five detected in X-rays, as well, see
Table~\ref{tab:source_info}). Hereafter, they will be referred to as
x$\textunderscore$NLSy1 galaxies and
g$\textunderscore$NLSy1 galaxies, respectively. Their
  B-magnitude distributions are shown in Fig.~\ref{apparent B-band} (upper panel for the x$\textunderscore$NLSy1
  and bottom panel for the g$\textunderscore$NLSy1 galaxies).

\begin{table*}
 \centering
 \begin{minipage}{400mm}
 {\small
   \caption[caption]{Observational details and the INOV status inferred for the sets of 18 x$\textunderscore$NLSy1 and 7 g$\textunderscore$NLSy1 galaxies monitored in total 45 sessions \\\hspace{\textwidth}(photometric aperture radius used = 2$\times$FWHM).}
\label{NLSy1:tab_result}
 \begin{tabular}{ccc ccccc ccr}\\
   \hline
   {NLSy1} &{Date$^{a}$} &  {T$^{b}$}  & {N$^{c}$}  & Median$^{d}$ & { $F^{\eta}$-test} & {INOV } & { $F_{enh}$-test} & {INOV} &{$\sqrt { \langle \sigma^2_{i,err} \rangle}$} & $\psi^\dagger$\\
   (SDSS name) & yyyy.mm.dd &  & & FWHM  & {$F_1^{\eta}$},{$F_2^{\eta}$} & status$^{e}$ & $F_{enh}$ & status$^{f}$  & (AGN-s)$^\dagger$ & (\%) \\
   && (hr) && (arcsec) &           &{99\%}& & {99\%}&&\\
   {(1)}&{(2)} & {(3)} & {(4)} & {(5)} & {(6)} & {(7)} & {(8)} & {(9)} & {(10)} & {(11)} \\
   \hline
   \multicolumn{10}{| c |}{\it x$\textunderscore$NLSy1s} \\\hline
 J071340.30$+$382039.8 & 2016.03.10 & 4.34& 54& 1.98& 0.23, 0.19  &  NV , NV &  1.39 &  NV& 0.020 & --\\
 J073623.14$+$392617.9 & 2015.11.04 & 4.30& 45& 2.10& 0.31, 0.11  &  NV , NV &  1.40 &  NV& 0.015 & --\\                             
 J075245.60$+$261735.9 & 2016.11.22 & 4.32& 31& 2.30& 0.57, 0.43  &  NV , NV &  0.81 &  NV& 0.006 & --\\                                 
 J080638.98$+$724820.5 & 2016.02.06 & 4.32& 40& 3.93& 0.19, 0.19  &  NV , NV &  0.51 &  NV& 0.017 & --\\
                       & 2017.11.19 & 5.22& 41& 1.86& 2.77, 3.28  &   V , V  &  4.12 &  V & 0.006 & 3.76\\
                       & 2019.02.03 & 3.60& 32& 2.18& 14.01, 20.46 &  V , V  &  2.80 &  V & 0.008 & 13.29\\
 J093703.02$+$361537.1 & 2016.11.23 & 3.47& 28& 2.60& 0.45, 0.50  &  NV , NV &  0.97 &  NV& 0.010 & --\\
 J100541.85$+$433240.2 & 2016.12.01 & 4.00& 37& 1.86& 0.42, 0.56  &  NV , NV &  0.65 &  NV& 0.007 & --\\
 J101000.70$+$300321.6 & 2016.11.30 & 2.94& 27& 2.26& 0.40, 0.47  &  NV , NV &  0.83 &  NV& 0.008 & --\\
                       & 2019.03.10 & 3.56& 30& 2.11& 0.55, 0.46  &  NV , NV &  0.83 &  NV& 0.012 & --\\
 J140516.22$+$255533.9 & 2019.03.23 & 3.88& 26& 2.39& 0.81, 0.92  &  NV , NV &  0.64 &  NV& 0.016 & --\\
  J140827.81$+$240924.8 & 2019.03.26 & 3.21& 32& 3.17& 0.62, 0.68  &  NV , NV &  1.20 &  NV& 0.012 & --\\
 J144240.80$+$262332.6 & 2017.05.05 & 3.90& 23& 1.86& 0.74, 0.55  &  NV , NV &  2.08 &  NV& 0.010 & --\\                        
 J144825.10$+$355946.7 & 2017.05.04 & 3.59& 39& 2.30& 0.35, 0.50  &  NV , NV &  0.83 &  NV& 0.010 & --\\
 J151936.15$+$283827.6 & 2019.02.16 & 3.33& 20& 3.07& 0.83, 0.81  &  NV , NV &  1.14 &  NV& 0.010 & --\\
 J162901.32$+$400759.5 & 2017.05.05 & 3.30& 24& 2.14& 0.66, 0.65  &  NV , NV &  2.54 &  V & 0.020 & 5.60\\
                       & 2019.04.09 & 3.13& 22& 2.35& 0.77, 0.47  &  NV , NV &  0.95 &  NV& 0.017 & -- \\ 
 J163323.59$+$471859.0 & 2017.05.20 & 4.33& 36& 2.26& 0.93, 0.76  &  NV , NV &  1.95 &  V & 0.007 & 2.44\\
                       & 2019.03.20 & 3.69& 33& 2.66& 1.68, 1.78  &  NV , NV &  2.15 &  V & 0.016 & 9.16\\
  J170231.05$+$324719.7 & 2019.03.10 & 3.01& 26& 2.18& 0.88, 0.64  &  NV , NV &  0.87 &  NV& 0.006 & --\\
 J170330.38$+$454047.3 & 2017.06.03 & 3.76& 37& 2.41& 0.51, 0.75  &  NV , NV &  0.67 &  NV& 0.004 & --\\
                       & 2019.03.25 & 3.13& 45& 4.45& 0.63, 1.08  &  NV , NV &  0.59 &  NV& 0.010 & --\\
 J171304.46$+$352333.4 & 2017.05.04 & 3.35& 40& 2.11& 0.74, 0.55  &  NV , NV &  1.17 &  NV& 0.007 & --\\ 
 J171601.94$+$311213.7 & 2017.04.17 & 3.57& 29& 2.18& 0.48, 0.42  &  NV , NV &  1.63 &  NV& 0.005 & --\\ 
  \hline
                                    \multicolumn{10}{| c |}{\it g$\textunderscore$NLSy1s} \\\hline
 J032441.20$+$341045.0 & 2016.11.22 & 4.42 & 57& 2.32& 09.87, 13.87 &  V,  V   & 15.30 &   V  & 0.004 &  5.17\\  
                       & 2016.11.23 & 4.27 & 55& 2.13& 07.10, 12.24 &  V,  V   & 05.52 &   V  & 0.004 &  4.44\\
                       & 2016.12.02 & 4.41 & 47& 2.60& 89.82, 92.80 &  V,  V   & 57.04 &   V  & 0.003 & 11.36\\
                       & 2017.01.03 & 3.00 & 40& 2.47& 06.55, 10.03 &  V,  V   & 02.78 &   V  & 0.003 &  4.10\\
                       & 2017.01.04 & 3.39 & 36& 2.45& 33.28, 35.16 &  V,  V   & 12.58 &   V  & 0.003 &  7.52\\
  J084957.98$+$510829.0 & 2017.12.13 & 5.36 & 31& 2.83& 00.61, 00.68 &  NV, NV  & 00.81 &   NV & 0.034 & --\\
                       & 2019.04.08 & 3.04 & 13& 2.88& 00.65, 01.32 &  NV, NV  & 00.46 &   NV & 0.033 & --\\
 J094857.32$+$002225.6 & 2016.12.02 & 4.15 & 17& 2.58& 01.61, 01.76 &  NV, NV  & 05.99 &   V  & 0.017 &  7.82\\
                       & 2017.12.21 & 5.20 & 33& 2.24& 13.95, 16.31 &  V , V   & 26.48 &   V  & 0.011 & 16.49\\
 J110223.37$+$223920.5 & 2018.01.15 & 5.34 & 28& 2.66& 00.74, 00.92 &  NV, NV  & 02.99 &   V  & 0.018 &  4.87\\
                       & 2019.04.09 & 3.91 & 16& 2.51& 00.20, 00.31 &  NV, NV  & 00.82 &   NV & 0.017 &  --\\
 J122222.99$+$041315.9 & 2017.01.03 & 3.52 & 17& 2.38& 00.53, 00.25 &  NV, NV  & 00.73 &   NV & 0.020 & --\\  
                       & 2017.01.04 & 3.14 & 16& 2.36& 00.32, 00.37 &  NV, NV  & 01.95 &   NV & 0.014 & --\\ 
                       & 2017.02.21 & 4.44 & 41& 2.65& 00.74, 00.76 &  NV, NV  & 01.95 &   V  & 0.020 &  6.28\\ 
                       & 2017.02.22 & 5.50 & 51& 2.59& 03.98, 03.56 &  V , V   & 05.89 &   V  & 0.017 & 13.32\\ 
                       & 2017.03.04 & 4.93 & 40& 2.61& 00.73, 01.08 &  NV, NV  & 00.94 &   NV & 0.019 & --\\  
                       & 2017.03.24 & 3.94 & 39& 2.37& 00.93, 00.74 &  NV, NV  & 01.41 &   NV & 0.020 & --\\
 J150506.48$+$032630.8 & 2017.03.25 & 5.21 & 41& 2.08& 00.60, 00.60 &  NV, NV  & 01.70 &   NV & 0.028 & --\\
                       & 2018.04.12 & 3.05 & 19& 2.55& 00.67, 00.53 &  NV, NV  & 00.84 &   NV & 0.032 & --\\ 
 J164442.53$+$261913.3 & 2017.04.03 & 4.37 & 37& 2.50& 01.15, 01.44 &  NV, NV  & 03.23 &   V  & 0.012 &  5.66\\
                       & 2019.04.26 & 3.22 & 24& 2.27& 02.78, 03.06 &  V , V   & 03.75 &   V  & 0.011 &  7.38\\
                       
   \hline

    \multicolumn{10}{l}{{$^{a}$Date(s) of the monitoring session(s).}}\\
   \multicolumn{10}{l}{{$^{b}$Duration of the session.}}\\
   \multicolumn{10}{l}{{$^{c}$Number of data points in the DLCs of the monitoring session.}}\\
   \multicolumn{10}{l}{{$^{d}$Median FWHM  of the  seeing disk during the monitoring session.}}\\
    \multicolumn{10}{l}{$^{e}$V = Variable, if both DLCs show INOV at confidence level $>$ 99\% using the $F^{\eta}$-test, otherwise NV = Non-variable.}\\
   \multicolumn{10}{l}{$^{f}$V, if F$_{enh}>F_{c}(0.99)$ using the $F_{enh}$-test, otherwise NV.}\\ 
   \multicolumn{10}{l}{{$\dagger$Mean for the three DLCs of the AGN (i.e., relative to the three comparison stars).}}\\
    \end{tabular}  
 }              
 \end{minipage} 
    \end{table*}

\begin{table}
 
 {\small
   \caption{$\overline{DCs}$ and $\overline{\psi}$ of INOV computed for the subsets of NLSy1 galaxies.}
    \label{NLSy1:DC_result}
\centering
 \begin{tabular}{lrrrrr}
   \hline
   &  &\multicolumn{2}{| c |}{Using $F_{enh}$-test} & \multicolumn{2}{| c |}{Using $F^{\eta}$-test} \\
     \hline
         \multicolumn{1}{| c |}{NLSy1} & \multicolumn{1}{r}{Number} & \multicolumn{1}{| r |}{$^{\star}\overline{DC}$}   &{$^{\star}\overline{\psi}^{\dag}$} & {$^{\star}\overline{DC}$}   &  {$^{\star}\overline{\psi}^{\dag}$}  \\
               &  & ({\%}) &  ({\%})                  &  ({\%})  &     ({\%})             \\
  \multicolumn{1}{| c |}{(1)}          &   ({2})        &           ({3})          &     ({4})&    ({5})   &    ({6})     \\
        \hline
       $^{a}x\textunderscore$NLSy1 (RQ)       & 13 &   0     &   0     &    0     &    0       \\  
       $^{b}x\textunderscore$NLSy1 (RL)       &  5 &  43     &   7     &   13     &    8        \\
       $^{c}x\textunderscore$NLSy1 (RQ$+$RL)  & 18 &  12     &   7     &    4     &    8       \\
       $^{d}g\textunderscore$NLSy1 (RL)       &  7 &  53     &   8     &   31     &   11        \\
  \hline
  \multicolumn{6}{l}{$^{a}$For the 13 radio-quiet x$\textunderscore$NLSy1 galaxies.}\\
  \multicolumn{6}{l}{$^{b}$For the 5  radio-loud x$\textunderscore$NLSy1 galaxies.}\\
  \multicolumn{6}{l}{$^{c}$For the entire set of 18 (13 RQ$+$ 5 RL) x$\textunderscore$NLSy1 galaxies.}\\
  \multicolumn{6}{l}{$^{d}$For the entire set of 7 g$\textunderscore$NLSy1 galaxies (all radio-loud).}\\
  \multicolumn{6}{l}{$^{\star}$Using only good quality DLCs (24 for x$\textunderscore$NLSy1s and 21 }\\
  \multicolumn{6}{l}{for g$\textunderscore$NLSy1s) for which the mean error is $\lesssim$ 3\% (cf. Table~\ref{NLSy1:tab_result},}  \\
  \multicolumn{6}{l}{column 10) and the monitoring time is $\gtrsim$ 3.0 hours.}\\
  \multicolumn{6}{l}{$^{\dag}$This is the mean for all the DLCs belonging to the category `V'.}\\

  \end{tabular}  
     }         
     \end{table}

\section{Observations and Data Reduction}
\label{section3.0}
\subsection{Photometric observations}
All the 18 x$\textunderscore$NLSy1s and 7 g$\textunderscore$NLSy1s in our
sample were monitored in the Johnson-Cousin R (hereafter R$_{c}$)
filter, using the 1.3 metre (m) Devasthal Fast Optical Telescope
(DFOT) located at Devasthal near Nainital~
\citep{Sagar2010ASInC...1..203S} and the 1.04m Sampurnanand Telescope
(ST) ~\citep{Sagar1999CSci...77..643G} located at Nainital, both of
telescopes operated by the Aryabhatta Research Institute of
Observational Sciences (ARIES).  Out of the total 25 NLSy1
galaxies, 19 were monitored exclusively with the DFOT, 2 with the ST,
while the remaining 4 were monitored with both the telescopes. The
1.3m DFOT is a Ritchey-Chretien (RC) telescope with a fast beam (f/4)
and a pointing accuracy better than 10 arcsec rms.  It is equipped
with a 2k$\times$2k and a 512$\times$512 deep thermoelectrically
cooled (to about $-85^{\circ}$C) Andor CCD cameras. The 512$\times$512
CCD camera having a pixel size of 16 microns and a plate scale of 0.63
arcsec per pixel, covers a $\sim$ 5$\times$5 arcmin$^{2}$
field-of-view (FOV) on the sky. The CCD was read out at 1 MHz speed,
having rms noise and gain of 6.1 $e^-$ and 1.4 $e^-$ADU$^{-1}$,
respectively. The second camera on DFOT, with a 2k$\times$2k CCD, 
has a pixel size of 13.5 microns, a plate scale of 0.53 arcsec
per pixel and covers a FOV of $\sim$ 18$\times$18 arcmin$^{2}$ on the
sky. The 2k$\times$2k CCD was read out at 1 MHz speed, having a system
rms noise and gain of 7.5 $e^-$ and 2.0 $e^-$ ADU$^{-1}$,
respectively.  The 1.04m ST is also of the RC design, with an f/13
beam at the Cassegrain focus. The telescope is equipped with a
1k$\times$1k, a PyLoN 1340$\times$1300 and a 4k$\times$4k CCD
cameras. All three CCDs, viz., the 1k$\times$1k, the PyLoN
1340$\times$1300 and the 4k$\times$4k are cooled with Liquid nitrogen
(LN$_{2}$) to -120$^{\circ}$C. The 1k$\times$1k CCD (read noise = 7.0
$e^-$) has a gain of 11.98 $e^-$ ADU$^{-1}$ and with its pixel size of
24 microns, covers a FOV of 6.2$\times$6.2 arcmin$^{2}$ on the sky.
The PyLoN 1340$\times$1300 CCD camera with a pixel size of 20 microns
and a plate scale of 0.31 arcsec per pixel, covers a FOV of
6.8$\times$6.6 arcmin$^{2}$ on the sky.  Our monitoring with this CCD
camera was carried out at 1 MHz speed, a read noise of 6.4 $e^-$ and a
gain of 2 $e^-$ ADU$^{-1}$. The third camera, with a 4k$\times$4k CCD
chip, has a pixel size of 15 microns and a plate scale of 0.23 arcsec
per pixel, giving a FOV of $\sim$ 15$\times$15 arcmin$^{2}$ on the
sky. The CCD was set up at 4$\times$4 binning with a readout speed of
1 MHz and a gain of 3 $e^-$ ADU$^{-1}$ which corresponds to the read
noise of 10 $e^-$. As suggested by~\citet{Carini1990PhDT.......263C},
the probability of INOV detection can be significantly enhanced by
monitoring the target continuously for about 3-4 hours.  Accordingly,
we have monitored each of our target AGN (7 g$\textunderscore$NLSy1
and 18 x$\textunderscore$NLSy1 galaxies) continuously for $\geq$ 3.0
hours during each session. Typical exposure time for a session was set at between 4
and 15 minutes, depending on the brightness of the target source, the lunar
phase, and the sky transparency.

\subsection{Data reduction}
\label{sec3.2}
Pre-processing of the raw frames (bias subtraction, flat-fielding and
cosmic-ray removal) was done using the standard tasks within the {\textsc
  IRAF software package\footnote{Image Reduction and Analysis Facility
  (http://iraf.noao.edu/)}}.  Instrumental magnitudes of the monitored
NLSy1 and the three chosen comparison stars also recorded in the CCD image
frames were determined by aperture photometry~\citep
{1992ASPC...25..297S, 1987PASP...99..191S}, using the DAOPHOT II
algorithm\footnote{Dominion Astrophysical Observatory Photometry
  (http://www.astro.wisc.edu/sirtf/daophot2.pdf)}.  A crucial
parameter for the photometry is the radius of the photometric
aperture, which also determines the signal to noise ratio (SNR) of the
individual photometric data points. As suggested by~
\citet{Howell1989PASP..101..616H}, the maximum SNR of an object in a
frame is achieved when the photometric aperture radius is
approximately equal to the full width at half maximum (FWHM) of the
point spread function (PSF), and it decreases for both larger and
smaller apertures.  However, following the analysis
by~\citet{Cellone2000AJ....119.1534C}, the situation can be more
complex in the present situation where the targets are nearby and
hence the integrated light falling within the aperture centered on the
target AGN is often expected to have a significant contribution from
the underlying host galaxy.  Consequently, the relative contributions
from the AGN and the host galaxy within the aperture can vary
significantly with changing seeing disk during a session, and the
differential light curves (DLCs) may then show statistically significant,
albeit spurious variability. Taking note of this, we have first
determined for each CCD frame the seeing disk FWHM (by averaging
over the profiles of 5 bright but clearly unsaturated stars in that frame) and then
found the median FWHM for the entire session. The aperture photometry
for each session was then performed by setting the aperture radius
equal to 2$\times$FWHM, 3$\times$FWHM and 4$\times$FWHM. The resulting
DLCs of the target AGN were compared with the observed variation of
the seeing disk (FWHM) during the session, before proceeding with a
quantitative analysis. This has amounted to an extra-cautious approach,
since during our various monitoring sessions, FWHM remained mostly
between 2 to 3 arcsecs (except for two sessions) and hence an aperture
radius = 2$\times$FWHM is expected to yield fairly reliable DLCs for
the AGN even if its host galaxy is up to 2-mag brighter~\citep[see,
  table 2 of][]{Cellone2000AJ....119.1534C}.  For each monitored
NLSy1, we derived DLCs relative to three (steady) comparison stars
which we had selected on the basis of their proximity to the
target AGN, both in brightness and location within the CCD frame. In
our sample, except for 2 x$\textunderscore$NLSy1 galaxies, at least
one of the chosen comparison star (considered as reference star) falls within $\sim$ 1-mag of the
target NLSy1. Coordinates and other parameters of the reference/comparison stars
used for the different sessions are given in
Table~\ref{tab_xray_comp_star} and Table~\ref{tab_gray_comp_star} for
the x$\textunderscore$NLSy1 and g$\textunderscore$NLSy1 galaxies,
respectively. The g-r color difference for the target NLSy1 and the
corresponding chosen reference and comparison stars is always $<$ 0.8 and
$<$ 2.0, with the median values being 0.30 and 0.67, respectively
(column 7 in Table~\ref{tab_xray_comp_star} and
Table~\ref{tab_gray_comp_star}). It has been shown
by~\citet{Carini1992AJ....104...15C} and~\citet{2004MNRAS.350..175S}
that color differences of this order should produce a negligible effect
on the DLCs as the atmospheric attenuation changes during a monitoring
session.

\section{STATISTICAL ANALYSIS OF THE DLCs}
\label{sec4.0}
To check for the presence of INOV in a DLC, we have used two flavours of the 
\emph{F$-$test}: (i) the standard \emph{F$-$test}~\citep[hereafter $F^{\eta}-$test, see, e.g.,][]
{Goyal2012A&A...544A..37G} and (ii) the power-enhanced \emph{F$-$test}~
\citep[hereafter $F_{enh}-$test, see, e.g.,][]{Diego2014AJ....148...93D}. For the 
$F^{\eta}-$test, it is important that the rms errors on the photometric 
data points are corrected for the underestimation of the error in the 
instrumental magnitude returned by the routines in the standard softwares DAOPHOT and 
IRAF used here. It has been found in various studies that the error 
underestimation factor $\eta$ ranges between 1.3 and 1.75 
~\citep[][] {1995MNRAS.274..701G, 1999MNRAS.309..803G, 
Sagar2004MNRAS.348..176S, Stalin2004JApA...25....1S, Bachev2005MNRAS.358..774B}. 
Recently~\citet{Goyal2013JApA...34..273G} have estimated the best-fit value of 
$\eta$ to be 1.54$\pm$0.05, using the data acquired in 262 quasar monitoring 
sessions. 

The $F^{\eta}$-statistics can be written as~\citep[e.g.,][]{Goyal2012A&A...544A..37G}

\begin{equation} 
 \label{eq.fetest}
F_{1}^{\eta} = \frac{\sigma^{2}_{(q-s1)}} { \eta^2 \langle
  \sigma_{q-s1}^2 \rangle}, \nonumber \\
\hspace{0.2cm} F_{2}^{\eta} = \frac{\sigma^{2}_{(q-s2)}} { \eta^2
  \langle \sigma_{q-s2}^2 \rangle},\nonumber \\
\hspace{0.2cm} F_{s1-s2}^{\eta} = \frac{\sigma^{2}_{(s1-s2)}} { \eta^2
  \langle \sigma_{s1-s2}^2 \rangle}
\end{equation}
where $\sigma^{2}_{(q-s1)}$, $\sigma^{2}_{(q-s2)}$ and $\sigma^{2}_{(s1-s2)}$ are 
the variances of the `AGN-star1', `AGN-star2' and `star1-star2' DLCs, respectively, while
  $\langle \sigma_{q-s1}^2\rangle=\sum_ {i=1}^{N}\sigma^2_{i,err}(q-s1)/N$,  $\langle
\sigma_{q-s2}^2 \rangle$  and $\langle \sigma_{s1-s2}^2 \rangle$ are the mean square 
(formal) rms errors of the individual data points in the corresponding DLCs. As mentioned above, the scaling factor `$\eta$' is taken to be 
$1.5$ following~\citet{Goyal2013JApA...34..273G}.  \par
For $F^{\eta}$-test we have used the DLCs of `AGN-star1', `AGN-star2' and `star1-star2' displayed in 
the 2$^{nd}$, 3$^{rd}$ and 5$^{th}$ panels from the bottom in Figs.~1-6, respectively.

The $F$-values were computed for each DLC using Eq.~\ref{eq.fetest}, and
compared with the critical $F$ value, $F^{(\alpha)}_{\nu_{qs},\nu_{ss}}$, 
where $\alpha$ is the significance level set for the test, and $\nu_{qs}$ 
and $\nu_{ss}$ are the degrees of freedom for the `AGN-star' and 'star-star' 
DLCs (both degrees of freedom are equal to N-1  in the present case). Here, we set a critical 
significance level, $\alpha=$ 0.01 which corresponds to a confidence level of 
99\%, so that an AGN is designated as variable (V) if its computed 
$F$-value is found to be $>F_{c}(0.99)$ for both its DLCs (i.e., relative to the 
two comparison stars), or, non-variable (NV) if the computed $F$-value is found 
to be $<F_{c}(0.99)$ for either of the two DLCs. The computed $F^{\eta}$-values 
and the correspondingly inferred variability status for x$\textunderscore$NLSy1 and 
g$\textunderscore$NLSy1 galaxies are given in columns 6 and 7 of 
Table~\ref{NLSy1:tab_result}.\par

The second version of the \emph{F$-$test} employed here is the, so-called, 
$F_{enh}-$test~\citep[][]{Diego2014AJ....148...93D}. Its chief  merit
is that it transforms the DLCs of the comparison stars to the same 
photometric noise level as if the magnitudes of the comparison stars are 
exactly matched to the mean magnitude of the AGN monitored~\citep[thereby making 
the analysis free from the effect of the magnitude difference between the target
AGN and the comparison star(s), which can significantly impact some versions of the 
$F$-test wherein the factor $\eta$ is ignored, e.g., see][]{Joshi2011MNRAS.412.2717J}. 
Furthermore, the power of the $F_{enh}-$test increases with the number of comparison stars 
used. Therefore, in this work, for a given session of NLSy1 galaxy, we have chosen three (steady) comparison stars present in all the frames, among which the star with the closest match in magnitude to the target NLSy1 galaxy is taken as a reference star. The statistical criterion of $F_{enh}-$test is defined as:

\begin{equation}
\label{Fenh_eq}  
\hspace{1.0cm} F_{{\rm enh}} = \frac{s_{{\rm AGN}}^2}{s_{\rm c}^2}, 
\hspace{0.5cm} s_{\rm c}^2=\frac{1}{(\sum _{j=1}^k N_j)  - k }\sum _{j=1}^{k}
\sum _{i=1}^{N_j}s_{j,i}^2
\end{equation}

where $s_{{\rm AGN}}^2$ is the variance of the DLC of the target AGN 
(relative to the chosen reference star), while $s_{\rm c}^2$ is the stacked 
variance of the DLCs of the comparison stars and the reference star~
\citep{Diego2014AJ....148...93D}. $N_{j}$ is the number of observations 
of the $j^{th}$ star and $k$ is the total number of comparison stars used.\par
Here, $s_{{\rm j,i}}^2$ is the scaled square deviation defined as

\begin{equation}
\hspace{2.7cm} s_{j,i}^2=\omega _j(m_{j,i}-\bar{m}_{j})^2
\end{equation}

where $\omega_ {j}$ is the scaling factor, $m_{j,i}$'s are the differential 
instrumental magnitudes and $\bar{m_{j}}$ is the mean differential magnitude 
of the reference star and the $j^{th}$ comparison star.
Following~\citet{Joshi2011MNRAS.412.2717J}, we have taken $\omega_ {j}$ as 
the ratio of the averaged square error of the differential instrumental 
magnitudes in the `AGN-reference star' DLC to the averaged square error of 
the differential instrumental magnitudes in the `comparison star-reference star' 
DLC. For the DLC of the $j^{th}$ star

\begin{equation}
 \hspace{2.7cm} \omega _j=\frac{\langle\sigma^2_{i,err}(AGN-ref)\rangle}
 {\langle\sigma^2_{i,err}(s_{j}-ref)\rangle}
  \end{equation}

From the previous section, we may recall that the photometric errors returned by IRAF are underestimated by a factor of $\eta$=1.5. However, such underestimation will not affect the scaling factor $\omega_ {j}$ since it is designed to take care of any magnitude difference between the target AGN and the reference/comparison stars. Furthermore, by stacking variances of the comparison stars, the degrees of freedom of $F_{enh}-$test given in Eq.~\ref{Fenh_eq} can be increased and thereby the power of the test enhanced~\citep{Diego2014AJ....148...93D}.
For this test, we set the same critical significance level as we have used for the $F^{\eta}$-test.
Thus, we mark an AGN as variable (V) at 99\% confidence level, if the computed value of $F_{{\rm enh}}$
is found to be $> F_{c}(0.99)$  and non-variable (NV) otherwise.
The computed $F_{{\rm enh}}$ values (Eq.~\ref{Fenh_eq})
and the correspondingly assigned variability status for the sets of x$\textunderscore$NLSy1 and 
g$\textunderscore$NLSy1 galaxies are given in columns 8, 9 of Table~\ref{NLSy1:tab_result}.

\subsection{Estimation of duty cycle and amplitude of INOV}
Following the definition of INOV duty cycle (DC) given by~\citet{Romero1999A&AS..135..477R}, 
we have computed DC as

\begin{equation} 
\hspace{2.5cm} DC = 100\frac{\sum_{i=1}^n N_i(1/\Delta t_i)}{\sum_{i=1}^n (1/\Delta t_i)} 
\hspace{0.1cm}{\rm per cent} 
\label{eqno1} 
\end{equation} 

where $\Delta t_i = \Delta t_{i,obs}(1+$z$)^{-1}$ is the duration of the
$i^{th}$ monitoring session, corrected for the AGN's redshift, $z$. 
 The proposed definition of \textit{DC} takes into consideration the
actual monitoring duration $\Delta t{_i}$ for the $i^{th}$ session,
to compensate for the fact that the duration of the sessions for a
given AGN differs from night to night.
For $i^{th}$ session, $N_i$ is set equal to 1 if INOV is detected, otherwise $N_i$ is taken as 0.\par
To compute the average (i.e., representative) value of INOV duty cycle ($\overline{DC}$) for an AGN 
subclass/set, it is important to appreciate that the observational coverages of the different members of the
set are often unequal in terms of the overall duration of the monitoring sessions devoted to a given member.
Consequently, the computed DC of INOV for that AGN set would get biased in favour of the longer 
monitored member(s) of the set. For instance, the computed DC for an AGN set would be too high in 
case of longer overall monitoring durations were devoted in the campaign to its members which happen to be 
intrinsically more variable. In order to avoid this potential bias, we have first computed the INOV DC for  
individual members of the AGN subclass and then taken an average over all the members.
The computed average DC values of INOV for the different subclasses of  NLSy1s are listed in columns 3 and 5 of Table~\ref{NLSy1:DC_result}, based on the $F_{enh}$ and $F^{\eta}$-tests.\par

The amplitude of INOV ($\psi$) for the monitored AGN
is given by~\citep[e.g.,][]{Heidt1996A&A...305...42H}

\begin{equation} 
\hspace{2.5cm} \psi= \sqrt{({D_{max}}-{D_{min}})^2-2\sigma^2}
\end{equation} 

with $D_{min, max}$ = minimum (maximum) values in the AGN-star DLC
and $\sigma^2 = \eta^2\langle\sigma^2_{q-s}\rangle$, where,
$\langle\sigma^2_{q-s}\rangle$ is the mean square (formal) rms errors
of individual data points in the DLC and $\eta$
=1.5~\citep{Goyal2013JApA...34..273G}.\par

The mean value of the INOV amplitude ($\overline{\psi}$) for each AGN subclass (e.g., x$\textunderscore$NLSy1s and g$\textunderscore$NLSy1s) is computed by taking the average of $\psi$ over the DLCs with positive INOV detection (i.e., `V' category DLCs). The computed $\overline{\psi}$ values for the subclasses of NLSy1s covered in the present study are tabulated in columns 4 and 6 of Table~\ref{NLSy1:DC_result}, based on the $F_{enh}$ and $F^{\eta}$-tests.

\section{Results and discussion}
\label{section5.0}
To recapitulate, the two sets/subclasses of NLSy1 galaxies for which intranight
optical monitoring is reported here, are: x$\textunderscore$NLSy1
(detected in X-rays, but undetected in $\gamma$-rays) and
g$\textunderscore$NLSy1 (detected in $\gamma$-rays and, in some cases,
in X-rays as well). From Table~\ref{tab:source_info}, 18 galaxies belong
to the x$\textunderscore$NLSy1 subclass (out of which 5 are 
radio-loud) and 7 to the g$\textunderscore$NLSy1 subclass (all 7 being
radio-loud). We have monitored these two sets, respectively, in 24 and
21 sessions of a minimum 3-hour duration each. The computed mean
  values of the INOV duty cycle ($\overline{DC}$) for the two sets, based
on the $F_{enh}$-test and the $F^{\eta}$-test, at 99\% confidence
level (Sect.~\ref{sec4.0}), are given in
Table~\ref{NLSy1:DC_result}. We believe these estimates of the INOV
 duty cycle for NLSy1 galaxies are more representative compared to those reported
  in previous studies~\citep{Liu2010ApJ...715L.113L,
    Paliya2013MNRAS.428.2450P, Kshama2017MNRAS.466.2679K}. This
    assertion stems from the smaller likelihood of the INOV detections
    being spurious on account of any variations in the seeing disk
    during the monitoring session (see below and
    Sect.~\ref{sec3.2}). The reason for this is that, unlike the
  usual practice, we have taken into consideration the DLCs derived
  from aperture photometry employing not just one but 3 apertures sizes,
  having radii equal to 2, 3 and 4 times the median seeing (FWHM) for
  the session. This caution is specially important in case of the nearby
  ($z <$ 0.3) x$\textunderscore$NLSy1 and g$\textunderscore$NLSy1
  galaxies for which a (spurious) variation in the DLC, which
  correlates with the variation of the seeing disk can become
  statistically significant on account of a changing contribution from
  the host galaxy into the aperture during the monitoring session
  ~\citep[e.g., see][]{Cellone2000AJ....119.1534C}. From a careful
  inspection of the DLCs shown in Fig.~\ref{fig:lurve 3}, it is inferred
  that the impact of such a possibility can be significant for just one
  session (on 20$^{th}$ March 2019) when the (x$\textunderscore$NLSy1) galaxy
  J163323.59$+$471859.0 was monitored (see the top left panel of Fig.~\ref{fig:lurve 3}). However, since
  those DLCs exhibited highly significant variability even when
  aperture radii equal to 3 and 4 times the median seeing (FWHM) were
  used, the INOV detection claimed here is almost certainly genuine.\par
 
\begin{figure}
  \begin{minipage}[]{1.0\textwidth}
  \includegraphics[width=0.5\textwidth,height=0.27\textheight,angle=00]{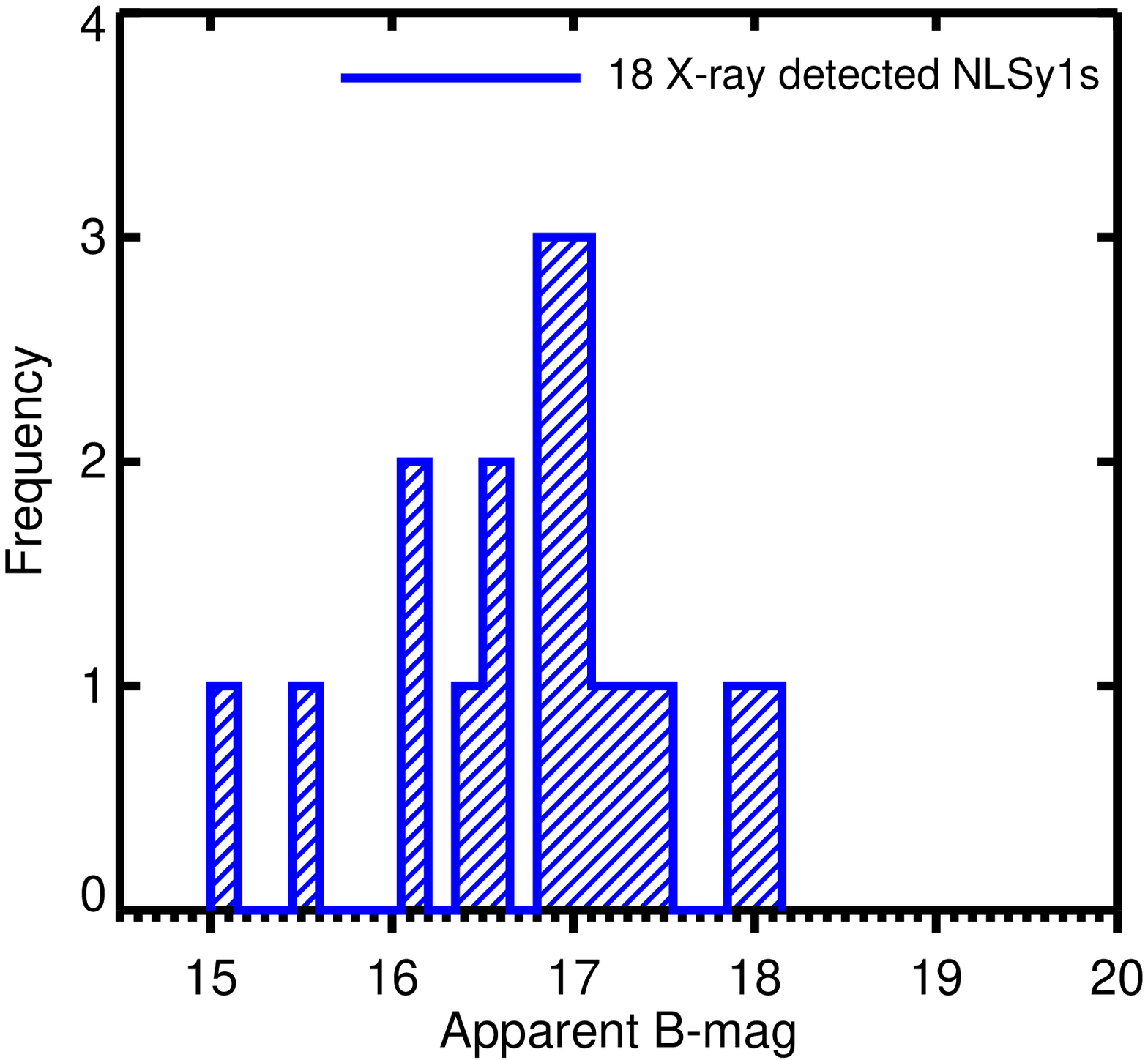} \\   
  \includegraphics[width=0.5\textwidth,height=0.27\textheight,angle=00]{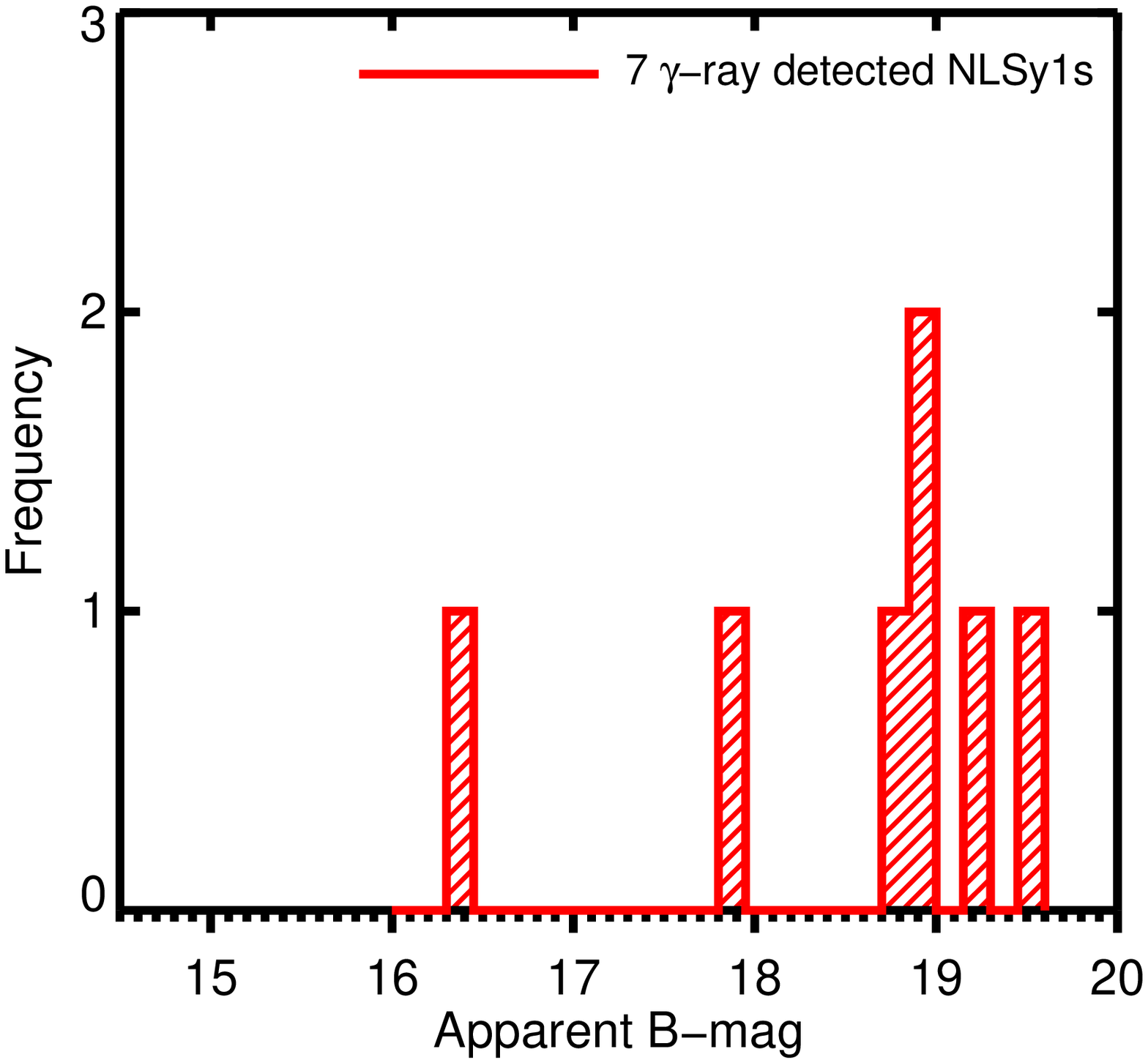}  
  \end{minipage}
  \caption{Distribution of B-band apparent magnitude for the 18 x$\textunderscore$NLSy1s (upper panel) and the 7 g$\textunderscore$NLSy1s (lower panel).}
  \label{apparent B-band}
\end{figure}

  It is seen that the
  INOV duty cycle (DC) for the x$\textunderscore$NLSy1s is 12\%,
  much smaller than the DC $\sim$53\% found here for the
  g$\textunderscore$NLSy1s. Conceivably, the difference might have
  stemmed from the mismatch between the limiting apparent magnitudes
  we adopted for selecting these two sets of NLSy1 galaxies
  (Sect.~\ref{section 2.0}). However, this possibility seems highly unlikely
  since the set of x$\textunderscore$NLSy1s should, in fact, be
  more amenable to INOV detection, being systemically brighter
  compared to the set of g$\textunderscore$NLSy1s (see
  Fig.~\ref{apparent B-band}). Considering now just the radio-loud subset of
  x$\textunderscore$NLSy1s (i.e., 5 out of the total 18 galaxies), it is seen
  from Table~\ref{NLSy1:DC_result} that they exhibit an INOV DC very
  similar to that found for the g$\textunderscore$NLSy1 galaxies (all
  7 of which are radio-loud). This point is further considered
  below.\par

As mentioned in Sect.~\ref{sec1.0}, detailed multi-band photometry
(combined with optical/near-infrared polarimetry) of the inner knots
in the jets of a few prominent blazars has strongly hinted that their
optical synchrotron emission consists of two physically distinct components, one 
linked to the X-ray synchrotron emission and the other to the radio
synchrotron emission~\citep{Jester2006ApJ...648..900J, Uchiyama2006ApJ...648..910U, Cara2013ApJ...773..186C}.
 One objective of the present study of NLSy1s was to investigate the INOV characteristics of
their optical synchrotron emission component linked to the X-ray emission, by carrying
out intranight monitoring of NLSy1s whose optical emission is more likely
to have a substantial contribution coming from the
X-ray linked optical synchrotron component. Clearly, such candidates
 are more likely to be picked up in X-ray detected samples of NLSy1s. 
It is then quite plausible that some such NLSy1s are present among 
the 13 radio-quiet members of the present set of 18 x$\textunderscore$NLSy1 
galaxies. The present finding of lack of INOV detection in all the 13 radio-quiet 
x$\textunderscore$NLSy1, in stark contrast to their 5 radio-loud counterparts, 
suggests that the INOV associated with the putative X-ray linked optical 
synchrotron component is rather muted.  Thus, it appears that the radio 
loudness level is the prime factor behind the INOV detection in NLSy1 galaxies and the pattern 
of the high-energy radiation plays only a minor role.
 A possible interpretation of this result is that, in relativistic jets, compared to their radio-emitting regions, the sites of 
synchrotron X-rays emission (and associated optical radiation) are more strongly 
magnetised and hence more stable against turbulence.\par

The high INOV duty cycle of $\sim$ 43\% estimated here for the radio-loud 
subset comprising of 5 x$\textunderscore$NLSy1s, compares well with 
the INOV DC of $\sim$ 53\% we find for the set of 7
g$\textunderscore$NLSy1s all of which are, as usual, radio-loud
(Table~\ref{tab:source_info}).  Interestingly, the similarity in INOV DC has 
been observed inspite of all the x$\textunderscore$NLSy1s in our
sample being undetected in gamma-rays (Sect.~\ref{section 2.0},
Table~\ref{tab:source_info}).  Here it may be recalled
that for quasars, neither radio-loudness nor radio core-dominance (a
marker of relativistic beaming) guarantees a pronounced INOV, only a
blazar characteristic, viz., a high optical polarisation  does~\citep{Goyal2012A&A...544A..37G,
Gopal-Krishna2018BSRSL..87..281G}. Thus, in the context of the present
finding, a polarimetric check on the blazar-like nature of the nuclei of the radio-loud 
x$\textunderscore$NLSy1 galaxies would be desirable.\par

It is possible to compare the present estimates of INOV DC for
x$\textunderscore$NLSy1 and g$\textunderscore$NLSy1 galaxies with
those reported for several other classes of (more luminous) AGNs
covered in the INOV study by~\citet{Goyal2013MNRAS.435.1300G}.  
Such a comparison is feasible because, like us, they too have applied the
$F^{\eta}$-test to the intranight DLCs and also set the confidence
level for INOV detection at 99\% (see Sect.~\ref{sec4.0}). Their
estimates of INOV duty cycles for INOV amplitude $\psi >$ 3\% are
$\sim$ 6, 11, 3, 10, 38 and 32 percent, respectively, for radio-quiet
quasars (RQQs), radio-intermediate quasars (RIQs), radio
lobe-dominated quasars (LDQs), low optical polarisation core-dominated
quasars (LPCDQs), high optical polarisation core-dominated quasars
(HPCDQs) and TeV blazars.  For our set of 7 g$\textunderscore$NLSy1s
galaxies (21 sessions), we find $\overline{DC}$ and $\overline{\psi}$ of $\sim$31\% and 11\%, respectively. 
As cited above, only blazars match such a high INOV duty cycle. In
fact, one might argue that the estimated INOV DCs of NLSy1 galaxies may 
be too low because of a substantial dilution of the optical synchrotron emission by
a comparatively steady thermal optical component which is contributed by the
host galaxy and the nuclear accretion disk in these high Eddington accretors
\citep[Sect.~\ref{sec1.0}, also see,][]{Ojha2019MNRAS.483.3036O}. Further observational evidence is needed to assess this hypothesis.

\begin{figure*}
\centering
\epsfig{figure=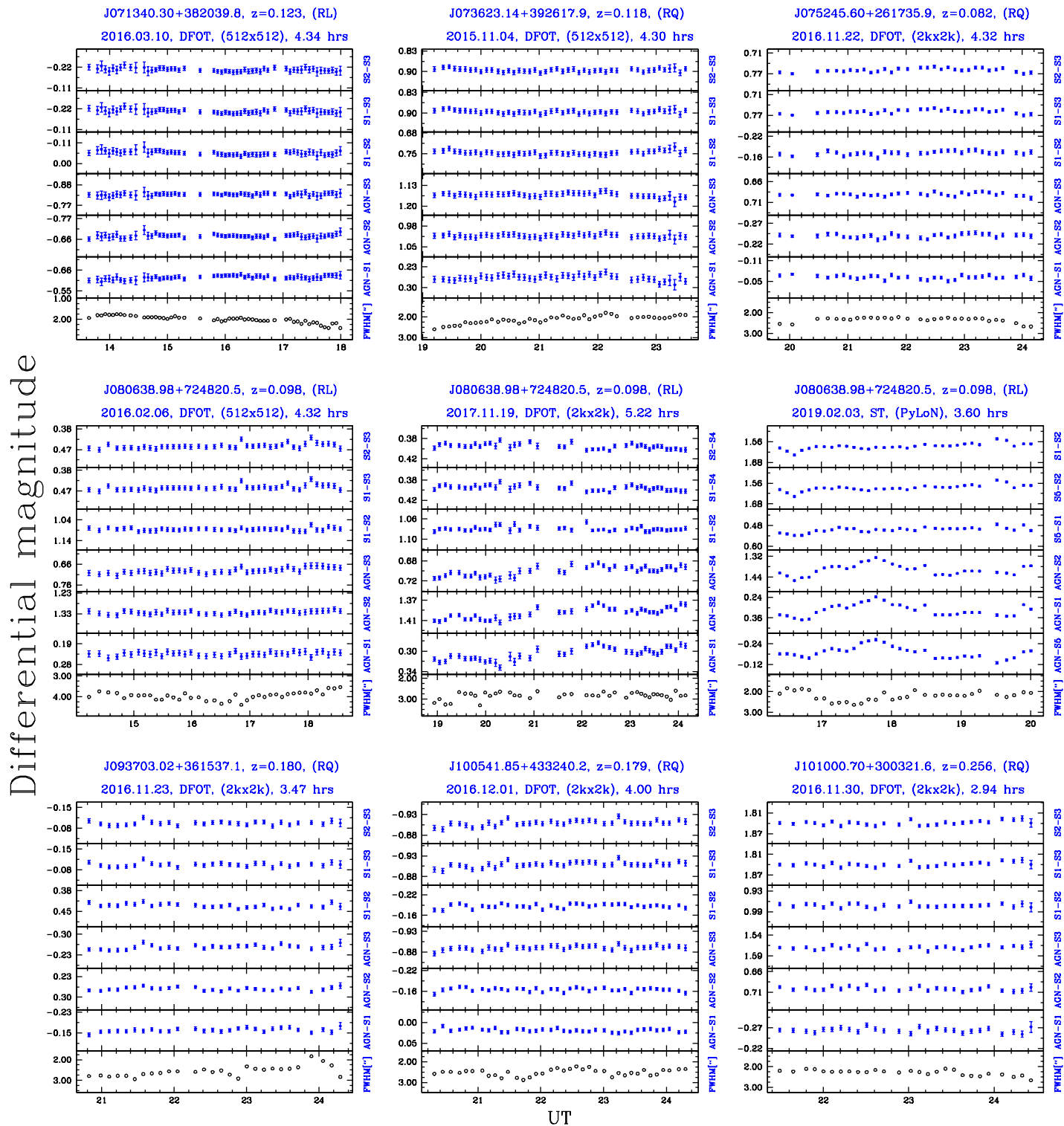}
\caption[]{Intranight differential light curves (DLCs) of the first 7
  x$\textunderscore$NLSy1s from our sample of 18
  x$\textunderscore$NLSy1s. The name, redshift, radio-classification
  and some observational details are given at the top of each
  panel. In each panel, the upper three DLCs are derived by pairing the
  chosen reference star (S1) and two comparison stars (S2 and S3), the lower three DLCs 
  are the `NLSy1-star' DLCs. The seeing disk (FWHM) variation during the session is displayed in the bottom panel, as defined in the labels on the right side.}
  
\label{fig:lurve 1}
\end{figure*}

\begin{figure*}
\centering
\epsfig{figure=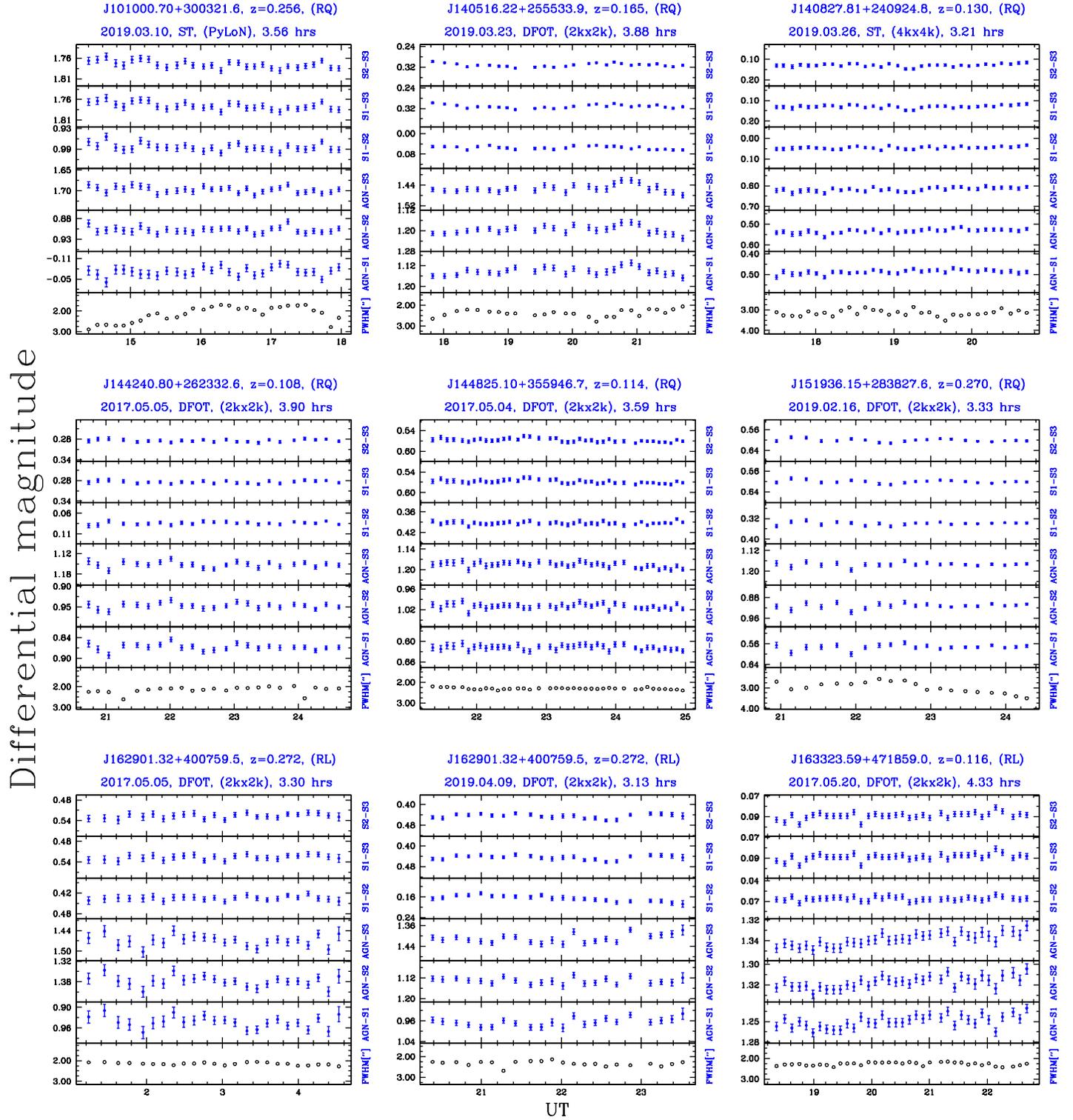}
\caption[]{Same as Fig.~\ref{fig:lurve 1}, but for the further 7 x$\textunderscore$NLSy1s from our 
sample of 18 x$\textunderscore$NLSy1s galaxies.}
\label{fig:lurve 2}
 \end{figure*}

\begin{figure*}
\epsfig{figure=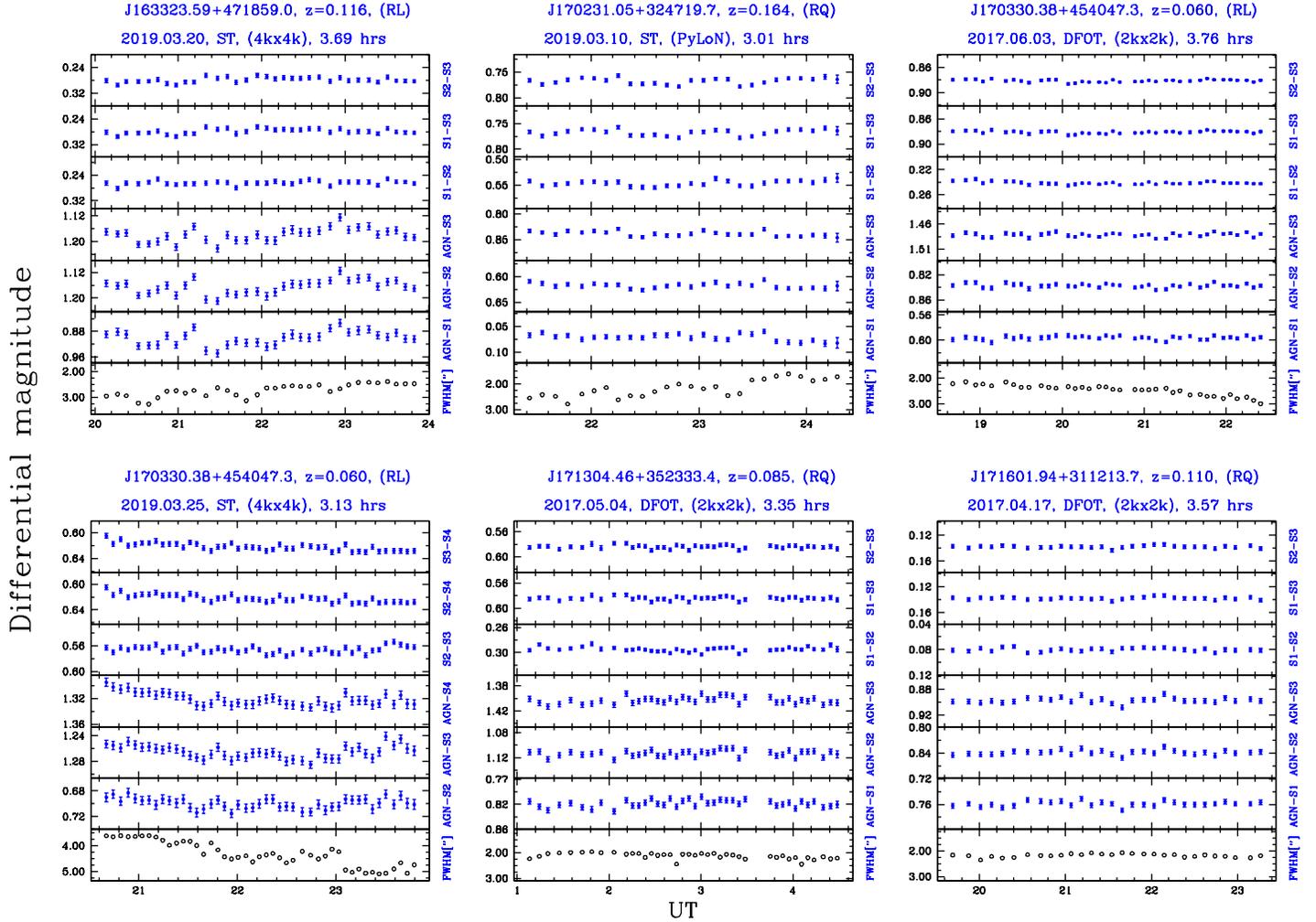,height=13.50cm,width=19.0cm,angle=0,bbllx= 18 bp,bblly=336 bp,bburx=559 bp,bbury=716 bp,clip=true}
\caption[]{Same as Fig.~\ref{fig:lurve 1}, but for the last 4 x$\textunderscore$NLSy1s from our sample of 18 x$\textunderscore$NLSy1s galaxies.}
\label{fig:lurve 3}
\end{figure*}

\begin{figure*}
\centering
\epsfig{figure=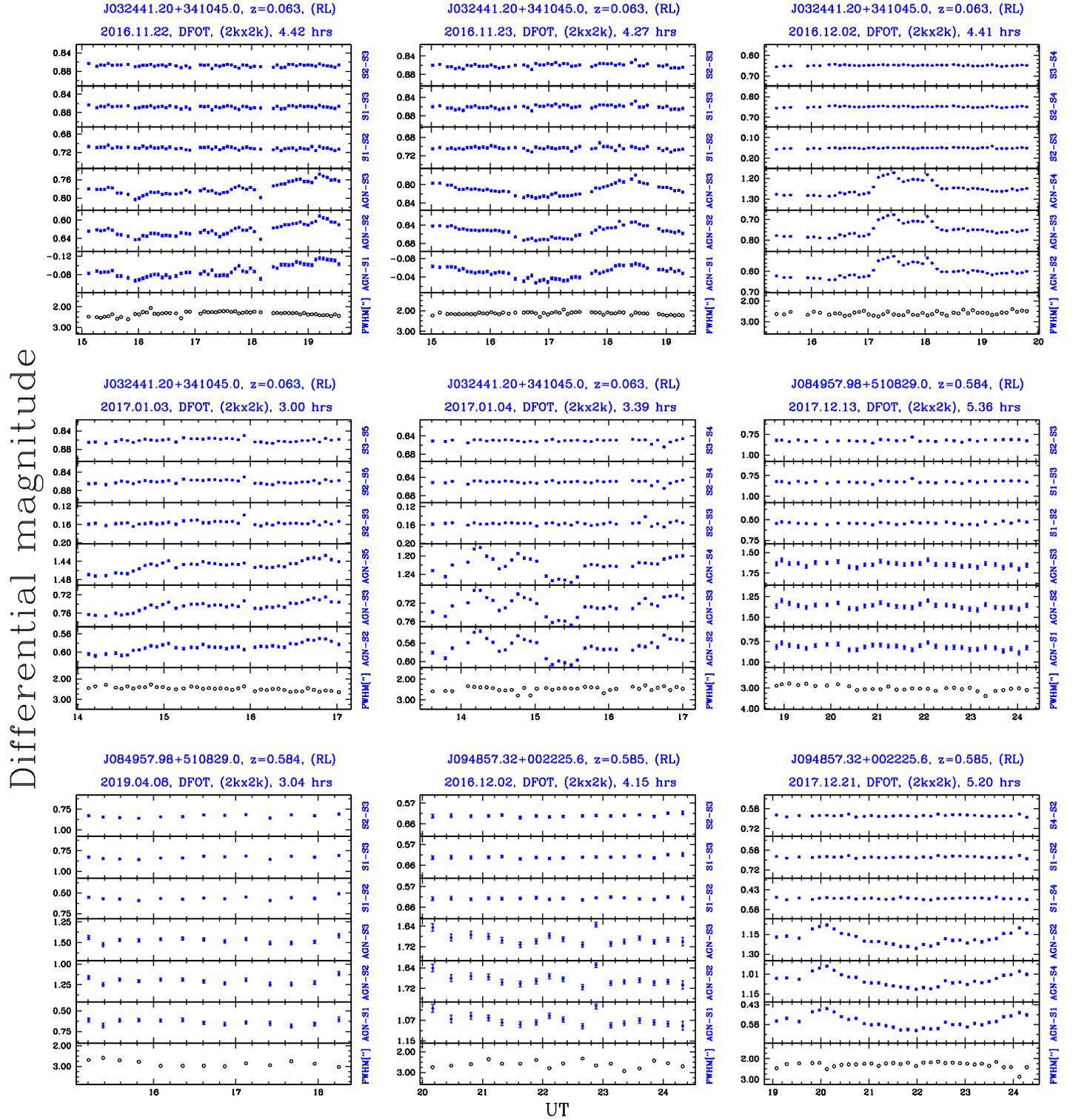}
\caption[]{Same as Fig.~\ref{fig:lurve 1}, but for 3 g$\textunderscore$NLSy1s from our sample of 7 g$\textunderscore$NLSy1s galaxies.}
\label{fig:lurve 5}
\end{figure*}

\begin{figure*}
\centering
\epsfig{figure=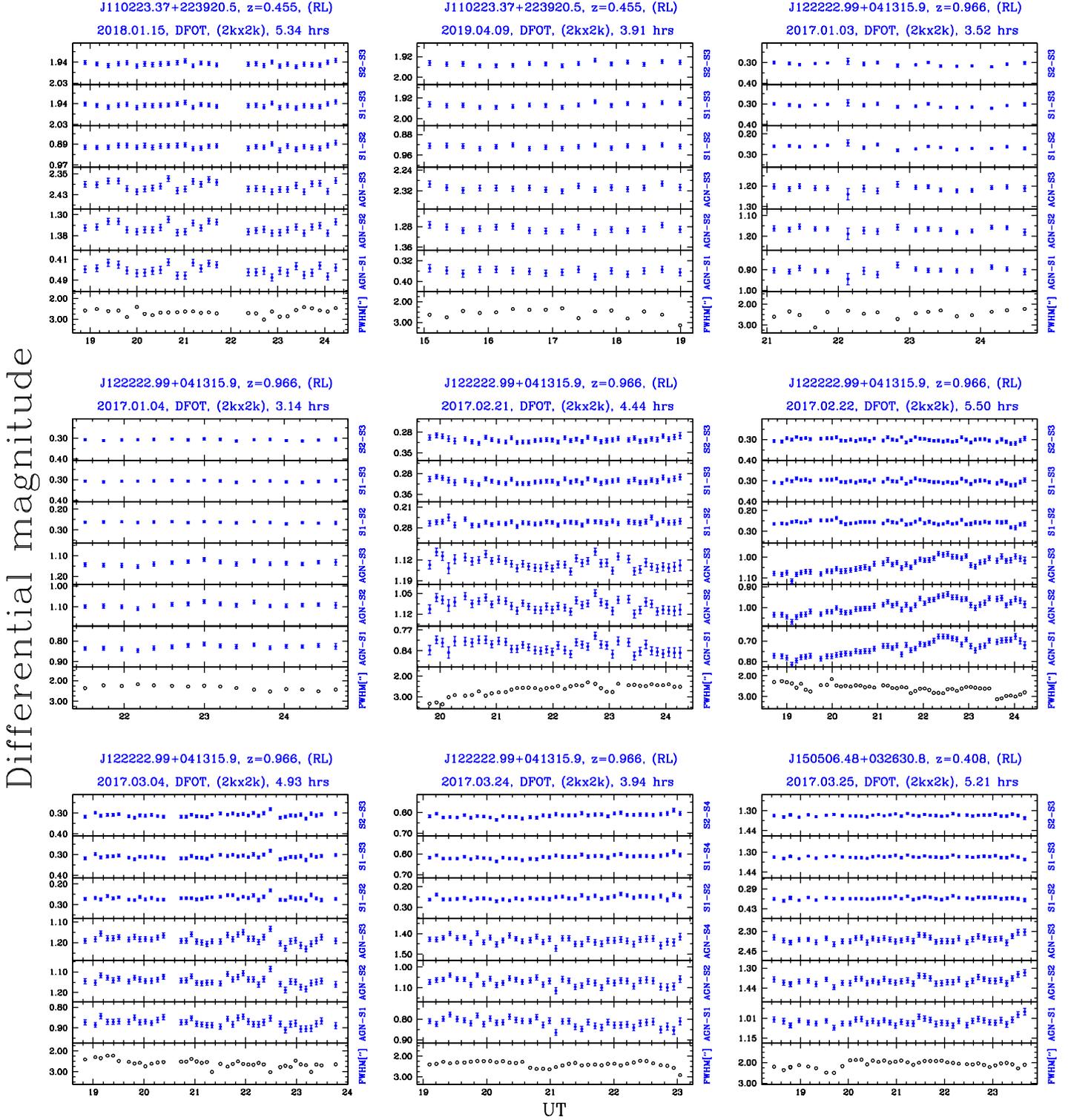}
\caption[]{Same as Fig.~\ref{fig:lurve 1}, but for the further 3 g$\textunderscore$NLSy1s from our sample of 7 g$\textunderscore$NLSy1s galaxies.}
\label{fig:lurve 6}
\end{figure*}

\begin{figure*}
\centering
\epsfig{figure=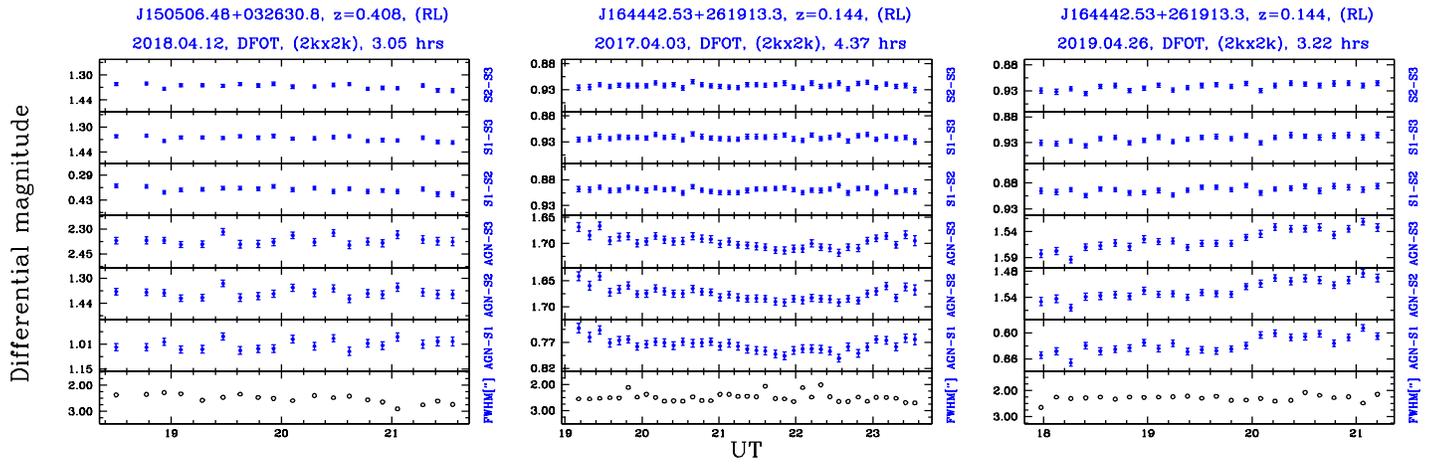,height=6.25cm,width=19.0cm,angle=0,bbllx= 18 bp,bblly=528 bp,bburx=559 bp,bbury=718 bp,clip=true}
\caption[]{Same as Fig~\ref{fig:lurve 1},  but for the last member of our sample of 7 g$\textunderscore$NLSy1s galaxies.}
\label{fig:lurve 7}
\end{figure*}

\section*{Acknowledgments}
{We thank an anonymous referee for his/her very important comments that helped us to improve this manuscript considerably.}

\bibliography{references}

\appendix

\section{Appendix}

\onecolumn
\begin{center}
\LTcapwidth=\textwidth  
\begin{longtable}{ccc ccc l}
\caption{The observational log and basic parameters of the comparison stars used for the 18 x$\textunderscore$NLSy1 galaxies.}
  \label{tab_xray_comp_star} \\

  \hline \multicolumn{1}{c}{{NLsy1}} & \multicolumn{1}{c}{{Date(s) of monitoring}} & \multicolumn{1}{c}{{R.A.(J2000)}}&\multicolumn{1}{c}{{Dec.(J2000)}} & \multicolumn{1}{c}{{\it g}} & \multicolumn{1}{c}{{\it r}} & \multicolumn{1}{c}{{\it g-r}}\\
\multicolumn{1}{c}{Comparison stars}          &      &   \multicolumn{1}{c}{(h m s)}       &\multicolumn{1}{c}{($^\circ$ $^\prime$ $^{\prime\prime}$)}   & \multicolumn{1}{c}{(mag)}   & \multicolumn{1}{c}{(mag)}   & \multicolumn{1}{c}{(mag)}     \\
 \multicolumn{1}{c}{(1)}      & \multicolumn{1}{c}{(2)}        & \multicolumn{1}{c}{(3)}           & \multicolumn{1}{c}{(4)}                              & \multicolumn{1}{c}{(5)}   & \multicolumn{1}{c}{(6)}   &  \multicolumn{1}{c}{(7)}     \\ \hline 
\endfirsthead

\multicolumn{7}{l}%
{{\tablename\ \thetable{} -- continued from previous page}} \\
\hline \multicolumn{1}{c}{{NLsy1}} & \multicolumn{1}{c}{{Date(s) of monitoring}} & \multicolumn{1}{c}{{R.A.(J2000)}}&\multicolumn{1}{c}{{Dec.(J2000)}} & \multicolumn{1}{c}{{\it g}} & \multicolumn{1}{c}{{\it r}} & \multicolumn{1}{c}{{\it g-r}}\\
\multicolumn{1}{c}{Comparison stars}          &      &   \multicolumn{1}{c}{(h m s)}       &\multicolumn{1}{c}{($^\circ$ $^\prime$ $^{\prime\prime}$)}   & \multicolumn{1}{c}{(mag)}   & \multicolumn{1}{c}{(mag)}   & \multicolumn{1}{c}{(mag)}     \\
 \multicolumn{1}{c}{(1)}      & \multicolumn{1}{c}{(2)}        & \multicolumn{1}{c}{(3)}           & \multicolumn{1}{c}{(4)}                              & \multicolumn{1}{c}{(5)}   & \multicolumn{1}{c}{(6)}   &  \multicolumn{1}{c}{(7)}     \\\hline 
\endhead

\hline \multicolumn{7}{r}{{Continued on next page}} \\

\endfoot

\endlastfoot

J071340.30$+$382039.8 & 2016 March 10  & 07 13 40.29 & $+$38 20 39.83 & 15.40 & 15.00  & 0.40$^{*}$         \\
S1                    &                & 07 13 45.41 & $+$38 20 08.52 & 17.40 & 15.40  & 2.00$^{*}$         \\
S2                    &                & 07 13 47.42 & $+$38 21 36.36 & 16.50 & 15.40  & 1.10$^{*}$         \\	 
S3                    &                & 07 13 41.90 & $+$38 20 57.14 & 16.00 & 15.30  & 0.70$^{*}$         \\
J073623.14$+$392617.9 & 2015 Nov. 04   & 07 36 23.13 & $+$39 26 17.88 & 16.22 & 15.91  & 0.31     \\		 
S1                    &                & 07 36 27.30 & $+$39 28 36.67 & 16.12 & 15.56  & 0.56         \\
S2                    &                & 07 36 11.18 & $+$39 26 29.61 & 15.23 & 14.79  & 0.44         \\	 
S3                    &                & 07 36 22.37 & $+$39 27 17.39 & 15.20 & 15.16  & 0.04         \\	 	 
J075245.60$+$261735.9 & 2016 Nov. 22   & 07 52 45.60 & $+$26 17 35.88 & 16.79 & 16.65  & 0.14     \\	 
S1                    &                & 07 53 03.29 & $+$26 17 08.70 & 16.93 & 16.52  & 0.41         \\	 
S2                    &                & 07 52 49.56 & $+$26 17 07.74 & 17.01 & 16.67  & 0.34         \\
S3                    &                & 07 52 49.31 & $+$26 17 36.11 & 16.07 & 15.74  & 0.33         \\
J080638.98$+$724820.5 & 2016 Feb. 06, 2017 Nov. 19, 2019 Feb. 03  & 08 06 38.98 & $+$72 48 20.53 & 16.40 & 15.80  & 0.60$^{*}$     \\	 
S1                    & 2016 Feb. 06, 2017 Nov. 19, 2019 Feb. 03  & 08 06 20.59 & $+$72 48 16.00 & 15.74 & 14.23  & 1.44$^{*}$      \\	 
S2                    & 2016 Feb. 06, 2017 Nov. 19, 2019 Feb. 03  & 08 06 22.31 & $+$72 48 47.69 & 15.50 & 14.70  & 0.80$^{*}$       \\
S3                    & 2016 Feb. 06                              & 08 06 25.19 & $+$72 49 18.15 & 17.00 & 15.60  & 1.40$^{*}$      \\	 
S4                    & 2017 Nov. 19                              & 08 06 14.47 & $+$72 51 48.14 & 16.40 & 15.40  & 1.00$^{*}$         \\
S5                    & 2019 Feb. 03                              & 08 07 17.42 & $+$72 48 41.73 & 17.20 & 16.20  & 1.00$^{*}$         \\
J093703.02$+$361537.1 & 2016 Nov. 23 & 09 37 03.02 & $+$36 15 37.08 & 17.58 & 17.00  & 0.58      \\
S1                    &              & 09 37 01.85 & $+$36 14 47.64 & 18.14 & 17.14  & 1.00     \\	 
S2                    &              & 09 37 21.41 & $+$36 19 18.33 & 18.25 & 16.80  & 1.45         \\
S3                    &              & 09 36 30.15 & $+$36 21 34.08 & 17.67 & 17.23  & 0.44         \\
J100541.85$+$433240.2 & 2016 Dec. 01 & 10 05 41.85 & $+$43 32 40.19 & 16.39 & 16.37  & 0.02     \\
S1                    &              & 10 05 31.35 & $+$43 24 21.76 & 17.57 & 16.58  & 0.99         \\
S2                    &              & 10 05 32.95 & $+$43 30 34.54 & 17.77 & 16.71  & 1.06         \\
S3                    &              & 10 05 28.77 & $+$43 32 38.09 & 18.45 & 17.43  & 1.02         \\
J101000.70$+$300321.6 & 2016 Nov. 30, 2019 March 10 & 10 10 00.70 & $+$30 03 21.60 & 16.97 & 16.82  & 0.15     \\
S1                    &              & 10 10 12.91 & $+$30 04 34.61 & 18.28 & 17.13  & 1.15         \\
S2                    &              & 10 10 10.51 & $+$30 02 25.28 & 16.65 & 16.11  & 0.54         \\
S3                    &              & 10 09 48.18 & $+$30 03 02.45 & 15.64 & 15.24  & 0.40         \\
J140516.22$+$255533.9 & 2019 March 23& 14 05 16.22 & $+$25 55 33.96 & 15.29 & 15.40  & \hspace{-0.2 cm}$-$0.11  \\
S1                    &              & 14 05 10.82 & $+$25 58 08.37 & 14.97 & 14.49  & 0.48         \\
S2                    &              & 14 05 18.50 & $+$26 01 43.42 & 16.02 & 14.87  & 1.15         \\	 
S3                    &              & 14 05 12.93 & $+$26 03 51.49 & 16.20 & 14.89  & 1.31         \\
J140827.81$+$240924.8 & 2019 March 26& 14 08 27.81 & $+$24 09 24.84 & 16.65 & 16.37  & 0.28     \\	 
S1                    &              & 14 07 59.37 & $+$24 15 02.86 & 16.18 & 15.71  & 0.47         \\	 
S2                    &              & 14 08 14.49 & $+$24 13 06.99 & 16.23 & 15.63  & 0.60         \\
S3                    &              & 14 08 05.74 & $+$24 10 52.96 & 16.69 & 15.78  & 0.91         \\
J144240.80$+$262332.6 & 2017 May 05  & 14 42 40.80 & $+$26 23 32.64 & 16.68 & 16.14  & 0.54      \\
S1                    &              & 14 41 55.52 & $+$26 20 19.87 & 16.03 & 15.50  & 0.53          \\	 
S2                    &              & 14 42 33.66 & $+$26 27 29.53 & 15.89 & 15.40  & 0.49          \\
S3                    &              & 14 41 49.35 & $+$26 19 26.46 & 15.65 & 15.21  & 0.44          \\
J144825.10$+$355946.7 & 2017 May 04  & 14 48 25.10 & $+$35 59 46.68 & 16.56 & 16.23  & 0.33      \\
S1                    &              & 14 48 04.84 & $+$35 56 47.01 & 17.01 & 15.69  & 0.32          \\	 
S2                    &              & 14 48 16.22 & $+$35 56 16.84 & 15.75 & 15.18  & 0.57          \\
S3                    &              & 14 48 09.86 & $+$36 04 22.75 & 15.38 & 14.98  & 0.40          \\
J151936.15$+$283827.6 & 2019 Feb. 16 & 15 19 36.14 & $+$28 38 27.67 & 17.05 & 16.80  & 0.25      \\
S1                    &              & 15 19 26.28 & $+$28 47 09 53 & 17.79 & 16.32  & 1.47          \\	 
S2                    &              & 15 19 12.21 & $+$28 41 45.49 & 16.96 & 15.74  & 1.22        \\
S3                    &              & 15 18 52.45 & $+$28 35 44.30 & 16.52 & 15.98  & 0.54        \\
J162901.32$+$400759.5 & 2017 May 05, 2019 April 09  & 16 29 01.32 & $+$40 07 59.53 & 17.77 & 16.64  & 0.13      \\
S1                    &              & 16 28 33.12 & $+$40 07 28.61 & 17.61 & 17.01  & 0.60          \\	 
S2                    &              & 16 28 29.55 & $+$40 05 26.72 & 18.13 & 16.74  & 1.39          \\
S3                    &              & 16 29 07.57 & $+$40 13 06.50 & 18.12 & 16.69  & 1.43          \\
J163323.59$+$471859.0 & 2017 May 20, 2019 March 20  & 16 33 23.59 & $+$47 18 59.04 & 17.24 & 16.94  & 0.30      \\       
S1                    &              & 16 32 59.26 & $+$47 26 05.45 & 15.57 & 15.18  & 0.39          \\	 
S2                    &              & 16 32 56.00 & $+$47 21 01.26 & 15.55 & 15.11  & 0.44          \\
S3                    &              & 16 33 10.21 & $+$47 14 56.91 & 15.42 & 15.07  & 0.35          \\ 
J170231.05$+$324719.7 & 2019 March 10 & 17 02 31.05 & $+$32 47 19.68 & 15.91 & 15.82  & 0.09      \\       
S1                    &              & 17 02 41.94 & $+$32 49 32.93 & 16.16 & 15.71  & 0.45          \\	 
S2                    &              & 17 02 32.16 & $+$32 49 17.32 & 15.77 & 15.18  & 0.59          \\
S3                    &              & 17 02 32.84 & $+$32 49 55.16 & 15.43 & 14.94  & 0.49          \\
J170330.38$+$454047.3 & 2017 June 03, 2019 March 25& 17 03 30.38 & $+$45 40 47.27 & 16.06 & 15.30  & 0.76      \\
S1                    & 2017 June 03               & 17 04 33.46 & $+$45 41 16.10 & 15.37 & 14.68  & 0.69          \\
S2                    & 2017 June 03, 2019 March 25& 17 04 02.02 & $+$45 42 16.56 & 15.02 & 14.39  & 0.63          \\	 
S3                    & 2017 June 03, 2019 March 25& 17 04 34.88 & $+$45 40 08.65 & 15.00 & 13.91  & 1.09          \\
S4                    & 2019 March 25              & 17 04 29.72 & $+$45 35 40.77 & 14.71 & 13.88  & 0.83          \\
J171304.46$+$352333.4 & 2017 May 04  & 17 13 04.46 & $+$35 23 33.36 & 16.27 & 16.03  & 0.24     \\
S1                    &              & 17 12 57.91 & $+$35 28 40.55 & 15.49 & 15.12  & 0.37          \\	 
S2                    &              & 17 13 33.48 & $+$35 19 13.05 & 15.52 & 15.08  & 0.44          \\	 
S3                    &              & 17 13 12.62 & $+$35 15 48.18 & 16.44 & 14.94  & 1.50          \\
J171601.94$+$311213.7 & 2017 April 17& 17 16 01.94 & $+$31 12 13.68 & 15.89 & 15.67  & 0.22      \\
S1                    &              & 17 15 47.58 & $+$31 08 22.95 & 15.31 & 14.81  & 0.50          \\	 
S2                    &              & 17 15 40.90 & $+$31 13 38.06 & 15.11 & 14.70  & 0.41          \\	 
S3                    &              & 17 15 42.64 & $+$31 07 13.85 & 15.15 & 14.67  & 0.48          \\

\hline
\multicolumn{7}{l}{ Position and apparent magnitude data have been taken from the SDSS DR14~\citep{Abolfathi2018ApJS..235...42A}.}\\
\multicolumn{7}{l}{$^{*}$Due to the unavailability of SDSS `g-r' colour, `B-R' colour has been used from the USNO-A2.0 catalog~\citep{Monet1998AAS...19312003M}.}\\
\end{longtable}
\end{center}

\begin{table*}
  \centering
 \resizebox{1.15\textwidth}{!}{\begin{minipage}{\textwidth}
 \caption{The observational log and basic parameters of the comparison stars used for the 7 g$\textunderscore$NLSy1 galaxies.}
    \label{tab_gray_comp_star}
\begin{tabular}{ccc ccc l}\\
\hline
{NLSy1} &   Date(s) of monitoring       &   {R.A.(J2000)} & {Dec.(J2000)}              & {\it g} & {\it r} & \hspace{0.1 cm} {\it g-r} \\
Comparison stars           &      &   (h m s)       &($^\circ$ $^\prime$ $^{\prime\prime}$)   & (mag)   & (mag)   & (mag)     \\
{(1)}      & {(2)}        & {(3)}           & {(4)}                              & {(5)}   & {(6)}   & \hspace{0.1 cm} {(7)}     \\
\hline
\multicolumn{7}{l}{}\\
J032441.20$+$341045.0 & 2016 Nov. 22, 23; Dec. 02; 2017 Jan. 03, 04& 03 24 41.20  &$+$34 10 45.00 & 14.50 & 13.70& 0.80$^{*}$   \\
S1                    & 2016 Nov. 22, 23                           & 03 24 46.40  &$+$34 06 37.57 & 16.00 & 15.20& 0.80$^{*}$    \\
S2                    & 2016 Nov. 22, 23; Dec. 02; 2017 Jan. 03, 04& 03 24 53.68  &$+$34 12 45.62 & 15.60 & 14.40& 1.20$^{*}$    \\
S3                    & 2016 Nov. 22, 23; Dec. 02; 2017 Jan. 03, 04& 03 24 53.55  &$+$34 11 16.58 & 16.20 & 14.40& 1.80$^{*}$    \\
S4                    & 2016 Dec. 02; 2017 Jan. 04                 & 03 25 05.53  &$+$34 17 57.26 & 14.80 & 14.20& 0.60$^{*}$    \\
S5                    & 2017 Jan. 03                               & 03 24 52.52  &$+$34 17 06.80 & 15.00 & 14.00& 1.00$^{*}$    \\
J084957.98$+$510829.0 & 2017 Dec. 13, 2019 April 08& 08 49 57.98  &$+$51 08 29.04 & 18.92 & 18.28& 0.64\\
S1                    &                           & 08 50 12.62  &$+$51 08 08.03 & 19.45 & 18.06& 1.39  \\
S2                    &                           & 08 50 39.07  &$+$51 04 59.77 & 17.95 & 17.37& 0.58  \\
S3                    &                           & 08 50 03.07  &$+$51 09 12.23 & 17.82 & 17.09& 0.73  \\
J094857.32$+$002225.6 & 2016 Dec. 02; 2017 Dec. 21& 09 48 57.32  &$+$00 22 25.56 & 18.59 & 18.43& 0.16    \\
S1                    & 2016 Dec. 02; 2017 Dec. 21& 09 48 36.95  &$+$00 24 22.55 & 17.69 & 17.28& 0.41    \\
S2                    & 2016 Dec. 02; 2017 Dec. 21& 09 48 37.47  &$+$00 20 37.02 & 17.79 & 16.70& 1.09    \\
S3                    & 2016 Dec. 02              & 09 48 53.69  &$+$00 24 54.91 & 17.52 & 16.65& 0.87    \\
S4                    & 2017 Dec. 21              & 09 49 00.44  &$+$00 22 34.91 & 18.27 & 16.84& 1.43    \\
J110223.37$+$223920.5 & 2018 Jan. 15, 2019 April 09& 11 02 23.37 &$+$22:47:14.20 & 19.57 & 18.10& 1.47    \\
S1                    &                           & 11 02 17.47  &$+$22 38 24.31 & 19.74 & 18.28& 1.46   \\
S2                    &                           & 11 02 57.30  &$+$22 46 05.04 & 17.58 & 17.06& 0.52    \\
S3                    &                           & 11 02 06.45  &$+$22 46 23.87 & 16.56 & 16.03& 0.53    \\
J122222.99$+$041315.9 & 2017 Jan. 03, 04; Feb. 21, 22; March 04, 24 & 12 22 22.99  &$+$04 13 15.95 & 17.02 & 16.80& 0.22  \\
S1                    & 2017 Jan. 03, 04; Feb. 21, 22; March 04, 24 & 12 22 34.02  &$+$04 13 21.57 & 18.63 & 17.19& 1.44      \\
S2                    & 2017 Jan. 03, 04; Feb. 21, 22; March 04, 24 & 12 21 56.12  &$+$04 15 15.19 & 17.22 & 16.78& 0.44    \\
S3                    & 2017 Jan. 03, 04; Feb. 21, 22; March 04     & 12 22 27.21  &$+$04 21 17.34 & 18.11 & 17.33& 0.78      \\
S4                    & 2017 March 24                               & 12 22 21.29  &$+$04 21 17.43 & 16.97 & 16.43& 0.54    \\
J150506.48$+$032630.8 & 2017 March 25; 2018 April 12& 15 05 06.48  &$+$03 26 30.84 & 18.64 & 18.22& 0.42    \\
S1                    &                              & 15 05 32.05  &$+$03 28 36.13 & 18.13 & 17.64& 0.49    \\
S2                    &                              & 15 05 16.38  &$+$03 28 07.52 & 18.76 & 17.34& 1.42    \\
S3                    &                              & 15 04 58.23  &$+$03 32 16.29 & 17.37 & 16.32& 1.05    \\
J164442.53$+$261913.3 & 2017 April 03; 2019 April 26 & 16 44 42.53  &$+$26 19 13.3  & 18.03 & 17.61& 0.42    \\
S1                    &                              & 16 44 54.56  &$+$26 23 20.38 & 18.14 & 16.79& 1.35    \\
S2                    &                              & 16 45 20.03  &$+$26 20 54.55 & 16.56 & 15.89& 0.67    \\
S3                    &                              & 16 44 34.40  &$+$26 15 30.27 & 16.28 & 15.80& 0.48    \\

\hline
\multicolumn{7}{l}{Position and apparent magnitude data have been taken from the SDSS DR14~\citep{Abolfathi2018ApJS..235...42A}.}\\
\multicolumn{7}{l}{$^{*}$Due to the unavailability of SDSS `g-r' colour, `B-R' colour has been used from the USNO-A2.0 catalog~\citep{Monet1998AAS...19312003M}.}\\
\end{tabular}
\end{minipage}}
\end{table*}

\label{lastpage}
 \end{document}